\documentclass[sigconf, nonacm]{acmart}

\usepackage{amsmath,amssymb,amsfonts}

\usepackage{xspace}
\usepackage{algorithmic}
\usepackage{algorithm}

\usepackage{multirow}
\usepackage{graphicx}
\usepackage{textcomp}
\usepackage{makecell}
\usepackage{pifont}
\usepackage{xcolor}
\usepackage[table,xcdraw]{xcolor}
\usepackage{enumitem}
\usepackage{tabularx}
\usepackage{tikz}
\usepackage{url}
\usepackage{booktabs}
\newcommand\vldbdoi{XX.XX/XXX.XX}
\newcommand\vldbpages{XXX-XXX}
\newcommand\vldbvolume{14}
\newcommand\vldbissue{1}
\newcommand\vldbyear{2020}
\newcommand\vldbauthors{\authors}
\newcommand\vldbtitle{\shorttitle} 
\newcommand\vldbavailabilityurl{https://anonymous.4open.science/r/lightCagra-F2D3}
\newcommand\vldbpagestyle{plain} 
\newcommand{\ourmethod}{\textsc{GRAB-ANNS}\xspace}

\newcommand{\cmark}{\textcolor{green!60!black}{\ding{51}}}   % ✓
\newcommand{\xmark}{\textcolor{red!70!black}{\ding{55}}}     % ✗
\newcolumntype{Y}{>{\centering\arraybackslash}X}

\begin{document}
\title{\ourmethod: High-Throughput Indexing and Hybrid Search via GPU-Native Bucketing}

%%
%% The "author" command and its associated commands are used to define the authors and their affiliations.
\author{Xinkui Zhao}
\affiliation{%
  \institution{Zhejiang University}
  \country{Hangzhou, China}
}
\email{zhaoxinkui@zju.edu.cn}

\author{Hengxuan Lou}
\affiliation{%
  \institution{Zhejiang University}
  \country{Hangzhou, China}
}
\email{12560049@zju.edu.cn}

\author{Yifan Zhang}
\affiliation{%
  \institution{Zhejiang University}
  \country{Hangzhou, China}
}
\email{12451018@zju.edu.cn}

\author{Junjie Dai}
\affiliation{%
  \institution{Zhejiang University}
  \country{Hangzhou, China}
}
\email{daijunjie@zju.edu.cn}

\author{Shuiguang Deng}
\affiliation{%
  \institution{Zhejiang University}
  \country{Hangzhou, China}
}
\email{dengsg@zju.edu.cn}

\author{Jianwei Yin}
\affiliation{%
  \institution{Zhejiang University}
  \country{Hangzhou, China}
}
\email{zjuyjw@cs.zju.edu.cn}


\begin{abstract}
Hybrid search, which jointly optimizes vector similarity and structured predicate filtering, has become a fundamental building block for modern AI-driven systems. While recent predicate-aware ANN indices improve filtering efficiency on CPUs, their performance is increasingly constrained by limited memory bandwidth and parallelism. Although GPUs offer massive parallelism and superior memory bandwidth, directly porting CPU-centric hybrid search algorithms to GPUs leads to severe performance degradation due to architectural mismatches, including irregular memory access, branch divergence, and excessive CPU–GPU synchronization.

In this paper, we present \ourmethod, a high-throughput, GPU-native graph index for dynamic hybrid search. Our key insight is to rethink hybrid indexing from a hardware-first perspective. We introduce a bucket-based memory layout that transforms range predicates into lightweight bucket selection, enabling coalesced memory accesses and efficient SIMT execution. To preserve global navigability under arbitrary filters, we design a hybrid graph topology that combines dense intra-bucket local edges with sparse inter-bucket remote edges. We further develop an append-only update pipeline that supports efficient batched insertions and parallel graph maintenance on GPUs.Extensive experiments on large-scale datasets show that \ourmethod achieves up to 240.1$\times$ higher query throughput and 12.6$\times$ faster index construction than state-of-the-art CPU-based systems, and up to 10$\times$ higher throughput compared to optimized GPU-native reimplementations, while maintaining high recall.

\end{abstract}

\maketitle

\pagestyle{\vldbpagestyle}
\begingroup\small\noindent\raggedright\textbf{PVLDB Reference Format:}\\
\vldbauthors. \vldbtitle. PVLDB, \vldbvolume(\vldbissue): \vldbpages, \vldbyear.\\
\href{https://doi.org/\vldbdoi}{doi:\vldbdoi}
\endgroup
\begingroup
\renewcommand\thefootnote{}\footnote{\noindent
This work is licensed under the Creative Commons BY-NC-ND 4.0 International License. Visit \url{https://creativecommons.org/licenses/by-nc-nd/4.0/} to view a copy of this license. For any use beyond those covered by this license, obtain permission by emailing \href{mailto:info@vldb.org}{info@vldb.org}. Copyright is held by the owner/author(s). Publication rights licensed to the VLDB Endowment. \\
\raggedright Proceedings of the VLDB Endowment, Vol. \vldbvolume, No. \vldbissue\ %
ISSN 2150-8097. \\
\href{https://doi.org/\vldbdoi}{doi:\vldbdoi} \\
}\addtocounter{footnote}{-1}\endgroup

\ifdefempty{\vldbavailabilityurl}{}{
\vspace{.3cm}
\begingroup\small\noindent\raggedright\textbf{PVLDB Artifact Availability:}\\
The source code, data, and/or other artifacts have been made available at \url{\vldbavailabilityurl}.
\endgroup
}

% \section{Introduction}
% \label{sec:introduction}
% About ANNS
%

%%%%%%%%%%%%5
% Paragraph 1: Background \& The Rise of Hybrid Search
% \begin{figure}
%     \centering
%     \includegraphics[width=1.0\linewidth]{figures/motivation-tmp.pdf}
%     \caption{The motivation of filter ANNS based on graph algorithm}
%     \label{fig:motivation}
% \end{figure}

\begin{table*}[t] % 使用 table* 让表格跨双栏，视觉冲击力更强
\centering
\caption{\textbf{Feature and Performance Landscape of State-of-the-Art Vector Indices.} \ourmethod uniquely bridges the gap between GPU acceleration and dynamic, range-aware flexibility, while excelling in high-dimensional datasets.}
\label{tab:comparison}
\resizebox{0.9\textwidth}{!}{% 调整宽度以适应页面
\begin{tabular}{l|c|ccc|cc}
\toprule
\multirow{2}{*}{\textbf{Method}} & \multirow{2}{*}{\textbf{Hardware}} & \multicolumn{3}{c|}{\textbf{Functionality}} & \multicolumn{2}{c}{\textbf{Performance}} \\ \cmidrule{3-7} 
 &  & \textbf{Range Filtering} & \textbf{Connectivity under Filter} & \textbf{Dynamic Update} & \textbf{Throughput} & \textbf{High-dim. Scalability} \\ 
\midrule
HNSW~\cite{malkov2018efficient}                    & CPU & \xmark & Low & \cmark & Low & Low \\
Filtered-DiskANN~\cite{10.1145/3543507.3583552} & CPU & \cmark & High & \cmark & Low & Low \\
SeRF~\cite{zuo2024serf}                       & CPU & \cmark & High & \cmark & Low & Low \\
ACORN~\cite{10.1145/3654923}                   & CPU & \cmark & High & \cmark & Low & Low \\
CAGRA~\cite{10597683}                 & GPU & \xmark & Low & \xmark & High & High \\
\midrule
\textbf{\ourmethod (Ours)}                      & \textbf{GPU} & \cmark & \textbf{High} & \textbf{\cmark} & \textbf{High} & \textbf{High} \\
\bottomrule
\end{tabular}%
}
\vspace{2pt}
% \footnotesize \emph{*High-Dim Scalability indicates performance stability as vector dimension grows (e.g., >1000d).}
\end{table*}

\section{Introduction}

Vector search has emerged as a cornerstone of modern information retrieval, underpinning a wide spectrum of latency-sensitive applications such as Retrieval-Augmented Generation (RAG)~\cite{arslan2024survey,fan2024survey,zhao2025trail}, agentic long-context memory~\cite{xu2025mem,long2025seeing,chhikara2025mem0}, and real-time recommendation~\cite{pan2024vector,rusum2024vector,sarwat2017database}. 
While Approximate Nearest Neighbor (ANN) techniques effectively retrieve semantically similar items, production workloads increasingly demand \textit{hybrid search}---the joint optimization of vector similarity with hard, structured predicates over metadata, such as time ranges or price limits. 
In practice, hybrid search is often implemented by loosely coupling a vector index with a separate metadata store. Under this architecture, vector similarity search and predicate filtering are executed by different subsystems, requiring complex integration logic and continuous data synchronization to ensure consistency~\cite{oceanbase_blog_seekdb,seekdb}. 

Within such loosely coupled designs, hybrid query execution is typically realized through either \textit{pre-filtering}~\cite{wang2021milvus,wei2020analyticdb} or \textit{post-filtering}~\cite{wang2021milvus,andoni2015practical,baranchuk2018revisiting,johnson2019billion,tencent2026cagra}. 
Pre-filtering first retrieves all records satisfying the structured predicate and then performs brute-force similarity search over the filtered subset. This approach scales poorly under medium-to-high selectivity predicates, as it effectively degenerates into large-scale linear scans. Conversely, post-filtering first performs ANN search over the entire vector index and subsequently removes candidates that fail the predicate. Since the nearest vectors may not satisfy the structured constraints, post-filtering often requires substantial over-retrieval to maintain recall, leading to unstable latency and unpredictable performance. 

To mitigate these limitations, recent works have developed \textit{predicate-aware} ANN indices that embed predicates directly into graph navigation. 
Approaches like Filtered-DiskANN~\cite{10.1145/3543507.3583552}, SeRF~\cite{zuo2024serf} and ACORN~\cite{10.1145/3654923} enhance query-time filtering through specialized structures such as stitched or segment graphs. 
However, as embedding dimensionality and query throughput demands continue to increase, hybrid search becomes increasingly constrained by the limited memory bandwidth and parallelism of traditional CPU architectures~\cite{zhao2025ame}.

Ideally, the massive parallelism and superior memory bandwidth of GPUs make them promising hardware candidates for overcoming these performance bottlenecks. However, directly porting state-of-the-art (SOTA) hybrid search algorithms to GPU architectures results in severe throughput degradation—up to 43$\times$ lower than their native CPU implementations (Figure~\ref{fig:motivation}a). Our profiling reveals that frequent CPU–GPU interactions dominate execution, introducing substantial synchronization overhead that accounts for up to 70\% of total latency and leaving actual GPU kernel execution to merely 20\% of the runtime (Figure~\ref{fig:motivation}b).

A deeper analysis attributes this inefficiency to a fundamental architectural mismatch. Existing hybrid search algorithms are inherently CPU-centric, relying on branch-heavy graph traversal and irregular pointer-chasing. Such characteristics conflict with the GPU's Single-Instruction Multiple-Thread (SIMT) execution model: irregular memory accesses hinder memory coalescing, while complex filtering logic causes severe warp divergence. As a result, GPU threads frequently serialize, leading to significant underutilization of computational resources.

To this end, we propose \ourmethod, a GPU-native hybrid search framework built on predicate-aware bucketing. 
\ourmethod bridges the gap between hardware-accelerated throughput and dynamic filtering flexibility, reconciling the trade-offs found in prior CPU-only or static GPU indices. 

From an algorithmic perspective, partitioning vectors into disjoint predicate ranges transforms expensive predicate filtering into lightweight bucket selection, effectively pruning the search space. 
Crucially, this logical partitioning translates into physical locality: by storing intra-bucket vectors contiguously, we enable coalesced memory accesses to exploit massive GPU parallelism. 
However, strict partitioning risks fragmenting the navigation graph and degrading global reachability.
GRAB-ANNS resolves this via a hybrid topology: \textit{local edges} provide dense intra-bucket connectivity for fast refinement, while sparse \textit{remote edges} bridge buckets to restore global navigability. 

To support dynamic requirements by AI workloads, \ourmethod employs an append-only pipeline: new vectors are ingested via lightweight indirection without data shifting, and the graph is updated in parallel via batched reverse rewiring with a freshness-biased pruning rule to fully employ the parallelism of GPU. 
At query time, we perform granularity-adaptive navigation: enforcing scalar pre-checks to skip invalid candidates, using remote edges for coarse-grained hops, and relying on local edges for fine-grained convergence within the valid manifold. 

Extensive experiments demonstrate that \ourmethod achieves up to 240.1$\times$ higher query throughput and 12.6$\times$ faster index construction time than state-of-the-art CPU-based hybrid search systems, while preserving high recall. Even compared with GPU-native reimplementations, \ourmethod still delivers up to 10$\times$ higher throughput.

Our key contributions are as follows:

\begin{itemize}[leftmargin=*]
    \item We propose \ourmethod, a high-throughput, GPU-native graph index for dynamic hybrid search. It rethinks hybrid indexing from a hardware-first perspective, reconciling predicate filtering flexibility with efficient SIMT-parallel execution on GPUs.
    
    \item We design a bucket-based memory layout that transforms range predicates into bucket selection, enabling lightweight batch insertions and coalesced memory accesses to fully exploit GPU parallelism.
    
    \item We introduce a remote-edge mechanism that complements intra-bucket local edges, restoring global graph connectivity under arbitrary range filters while preserving regular and highly parallel query execution.
    
    \item Extensive experiments show that \ourmethod achieves up to 240.1$\times$ higher query throughput and 12.6$\times$ faster index construction than state-of-the-art CPU-based hybrid search systems, and up to 10$\times$ higher throughput compared to optimized GPU-based implementations, while maintaining high recall.
\end{itemize}

    \section{Background}
% 2.1 算法基础：讲清楚 Small World 和连通性的重要性
\subsection{Graph-based ANNS}
\label{subsec:bg_graph}
Various indexing techniques have been proposed for ANNS, including tree-based~\cite{houle2014rank, lu2020vhp, muja2014scalable}, hashing-based~\cite{gong2020idec, jafari2020mmlsh, li2020efficient, lu2020r2lsh}, and quantization-based~\cite{guo2020accelerating, johnson2019billion} methods. 
Among these, graph-based~\cite{10597683, 8594636, malkov2014approximate, 9101583, 9739943, 9835618} indices typically offer the best trade-off between query latency and recall. 
These methods construct a proximity graph $G(V, E)$ where nodes represent data vectors and edges connect nearby neighbors. 

Query processing follows a Greedy Routing strategy, theoretically grounded in the \emph{Small World} property~\cite{malkov2014approximate}. 
To achieve efficient logarithmic scaling ($O(\log N)$), the graph topology must maintain both dense \emph{short-range links} for fine-grained local convergence and sparse \emph{long-range links} for rapid traversal across the vector space. 
However, as we will discuss, hard range filtering often severs these critical long-range pathways, fragmenting the graph into isolated islands and breaking global navigability.

% 2.2 硬件约束：讲清楚为什么 GPU 编程难（Coalescing, Divergence）
\subsection{GPU Hardware Architecture}
\label{subsec:bg_gpu}
To understand the design constraints of GPU-native indices, we highlight two critical architectural features of NVIDIA GPUs.

\textbf{Memory Hierarchy.}
NVIDIA GPUs utilize a hierarchical memory model. \textit{Device Memory} offers vast capacity but suffers from high latency; to maximize bandwidth, algorithms must ensure \textit{memory coalescing}, where consecutive threads access contiguous addresses. Conversely, \textit{Shared Memory} is a small, user-managed on-chip cache with extremely low latency, ideal for staging frequently accessed data.

\textbf{Thread Execution Model.}
Execution follows a Single-Instruction Multiple-Thread (SIMT) model, where threads are grouped into \textit{warps} (typically 32) that execute instructions in lock-step. Performance degrades significantly under \textit{Warp Divergence}, which occurs when threads follow different control paths (e.g., `if-else` branches), forcing the hardware to serialize execution. Consequently, GPUs favor regular, vector-like computations and are ill-suited for the irregular branching logic typical of tree-based indices.

% 2.3 现有方案：CAGRA 如何结合上述两者，以及它的死穴（静态 CSR）
\subsection{GPU-based Graph Index}
\label{subsec:bg_cagra}
To leverage the massive parallelism of modern hardware, CAGRA~\cite{10597683} has established itself as the state-of-the-art graph index on GPUs. 

\textbf{Static Construction Optimization.}
CAGRA employs a construction pipeline designed to saturate GPU throughput. It begins by initializing an approximate $k$-NN graph (e.g., via IVF-PQ) and refines it through rank-based pruning and reverse graph merging to enhance connectivity. 
Crucially, to satisfy the \emph{memory coalescing} requirement, the final graph is compacted into a static Compressed Sparse Row (CSR) format with a fixed degree per node.
While this rigid layout maximizes read bandwidth, it is inherently \textbf{hostile to dynamic updates}; inserting a single node requires shifting the entire data array, effectively forcing a full index reconstruction.

\textbf{Parallel Query Processing.}
For search, CAGRA adopts a cooperative execution model. Instead of relying on a single thread, it utilizes a warp to process a query in unison. 
The search initiates with randomly sampled seed nodes and iteratively explores the graph by fetching neighbors and computing distances in parallel. 
This approach mitigates warp divergence and ensures high-throughput execution for large-batch queries, provided the graph topology remains static.

% 3. 讲一下过滤的问题定义
\subsection{Problem Definition: Hybrid Search}

We consider a dataset $\mathcal{D} = \{ o_1, \dots, o_n \}$, where each object $o_i$ is a tuple $(v_i, s_i)$. Here, $v_i \in \mathbb{R}^d$ is a $d$-dimensional vector, and $s_i \in \mathbb{R}$ is the scalar predicate. A Range-Filtered ANNS query is defined as $Q = (q, k, [l, u])$, where $q$ is the query vector, $k$ is the target count, and $[l, u]$ denotes the filtering range.

\begin{definition}[Hybrid Search]
Given a query $Q = (q, k, [l, u])$, the goal is to retrieve a subset $\mathcal{R} \subset \mathcal{D}$ such that:
\begin{enumerate}
    \item $|\mathcal{R}| = k$;
    \item $\forall o_i \in \mathcal{R}$, the scalar predicate satisfies $l \le s_i \le u$;
    \item For any $o_i \in \mathcal{R}$ and $o_j \in \mathcal{D} \setminus \mathcal{R}$ satisfying $l \le s_j \le u$, we have $\text{dist}(q, v_i) \le \text{dist}(q, v_j)$.
\end{enumerate}
\end{definition}

Since exact search is often computationally prohibitive in high-dimensional spaces, we focus on the \textit{Approximate} Nearest Neighbor Search (ANNS). The search quality is typically evaluated by \textbf{Recall@$k$}. 
Let $\mathcal{R}^*$ denote the set of the true $k$ nearest neighbors of query $q$ that satisfy the filtering range $[l, u]$ (i.e., the ground truth). The recall is defined as the fraction of true neighbors successfully retrieved in the approximate result set $\mathcal{R}$:
\begin{equation}
    \text{Recall}@k = \frac{|\mathcal{R} \cap \mathcal{R}^*|}{k}
\end{equation}
A recall of 1.0 indicates that all the returned objects are the exact nearest neighbors within the specified range.

\begin{figure}[t]
    \centering
    \includegraphics[width=1.0\linewidth]{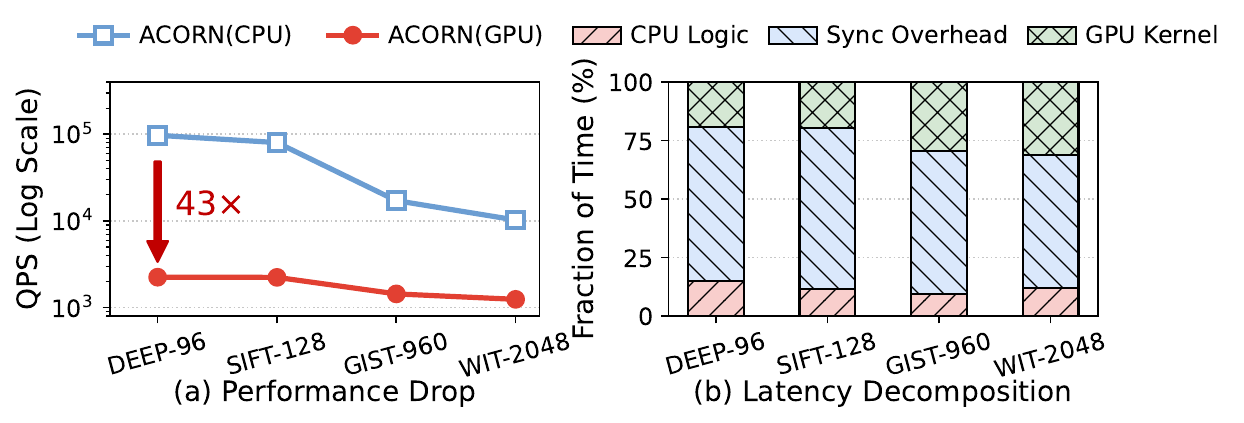}
    \caption{\textbf{Motivation.}(a) Query throughput (QPS) comparison between ACORN running on CPU and its GPU port across multiple datasets. (b) Latency breakdown of the GPU implementation.}
    \label{fig:motivation}
\end{figure}

\section{Motivation}
\label{sec:moti}
\textbf{Experiment: The Impact of GPUs on Hybrid Search.} Although numerous hybrid search methods have been proposed~\cite{10.1145/3654923,10.1145/3543507.3583552,zuo2024serf}, they are predominantly CPU-centric. To quantify the performance limitations of directly porting existing CPU-based hybrid search methods to GPU architectures, we conduct a microbenchmark using the state-of-the-art algorithm ACORN~\cite{10.1145/3654923} as a representative baseline. The evaluation is performed across four standard benchmarks (DEEP-96, SIFT-128, GIST-960, and WIT-2048) on an Intel Core i9-14900K CPU and an NVIDIA RTX 3090 Ti GPU. Detailed dataset configurations are provided in Section~\ref{sec:dataset}.

Figure~\ref{fig:motivation}a compares the query throughput (QPS) of the native CPU implementation against its direct GPU port across the different datasets. Contrary to expectations, utilizing the GPU does not improve performance; instead, it causes a severe throughput degradation of up to 43$\times$ on the DEEP dataset. To isolate the bottleneck of this inefficiency, we decompose the total query latency into three stages: CPU logic, CPU-GPU synchronization overhead, and GPU kernel execution. As illustrated in Figure~\ref{fig:motivation}b, the actual GPU kernel execution—where acceleration is expected—accounts for merely 20\% of the total latency. Ultimately, complex CPU-side control logic and the ensuing synchronization overhead heavily dominate the execution time.

\textbf{Analysis.} Based on the experimental results, we attribute this severe performance degradation to a fundamental architectural mismatch between CPU-centric hybrid search designs and the GPU execution model. Specifically, this mismatch manifests in two critical dimensions:

\textit{(1) Control Flow Conflict:} Existing hybrid search algorithms tightly couple vector similarity evaluations with complex, branch-intensive predicate filtering logics. While CPUs excel at handling complex control flows, GPUs operate on a Single-Instruction Multiple-Thread (SIMT) model~\cite{zhao2021survey,yi2024study}. Evaluating dynamic filter conditions during standard graph traversal forces threads within the same warp to take divergent execution paths. This severe warp divergence forces the hardware to serialize execution, leaving the massive parallel compute units of the GPU severely underutilized.

\textbf{(2) Memory Access Mismatch:} Standard graph traversal inherently relies on irregular, random memory access patterns (i.e., pointer-chasing) to fetch neighbors~\cite{malkov2014approximate}. CPUs mitigate this through large, complex cache hierarchies. Conversely, achieving high memory bandwidth on a GPU strictly requires coalesced memory accesses, where consecutive threads access contiguous memory addresses. The random node fetching in traditional filtered graphs completely breaks memory coalescing, creating a severe memory bandwidth bottleneck.

\section{Methodology}
\label{sec:methodology}

\begin{figure*}
    \centering
    \includegraphics[width=\textwidth]{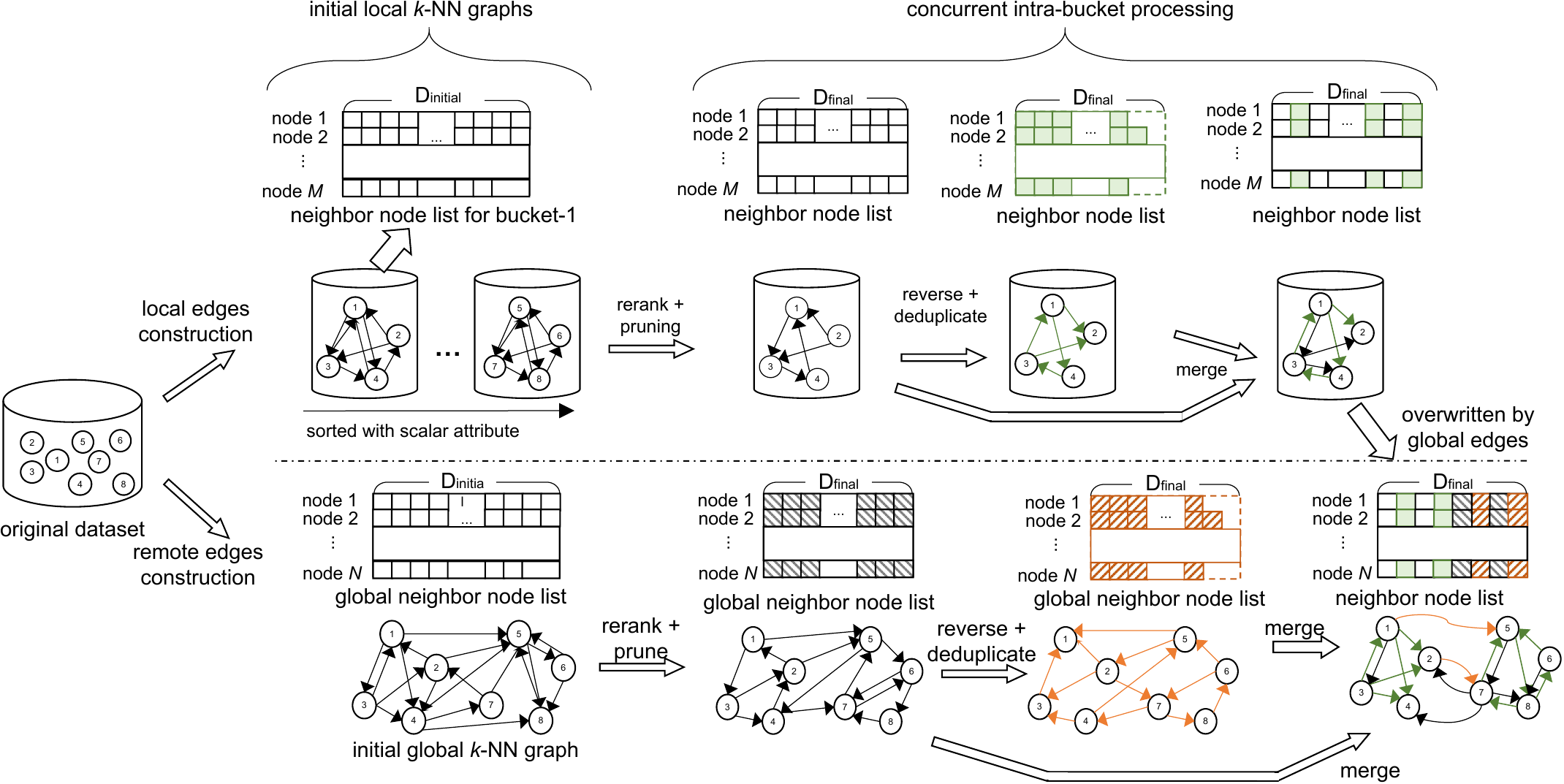}
    \caption{Construction flow of \ourmethod.}
    \label{fig:construction}
\end{figure*}

% ==========================================
% 4.1 Overview
% 目标：总览全局，一图胜千言
% ==========================================
\subsection{Overview}
\label{subsec:overview}
In this section, we present the overall architecture of \ourmethod, designed to deliver high-throughput range-filtered vector search on GPUs. Our approach consists of four key components:

\begin{itemize}[leftmargin=*]
    \item \textbf{Hardware-Aware Graph Topology:} To eliminate the irregular memory access patterns that plague traditional graph traversal, we partition data into ordered buckets based on scalar predicates and introduce a dual-layer edge structure (Local/Remote). Crucially, this topology enforces a strictly fixed node degree to guarantee coalesced memory accesses, while the remote edges restore global graph connectivity under arbitrary filter conditions.
    
    \item \textbf{Two-Phase Construction Strategy:} To saturate GPU compute units efficiently, we design a step-wise construction process. It first establishes dense intra-bucket local connectivity by leveraging highly parallel brute-force initializations optimized for small bucket sizes. Subsequently, it injects sparse inter-bucket remote links sampled from a global graph. This hardware-aligned strategy effectively balances local search precision with global navigability.
    
    \item \textbf{Dynamic Parallel Insertion:} Traditional static GPU graph layouts inherently resist dynamic updates due to cascading index shifts. To meet the real-time ingestion requirements of modern AI workloads, we propose a batch-oriented insertion algorithm that exploits massive GPU parallelism. It employs an append-only memory pipeline and parallel reverse rewiring to execute simultaneous updates of local and remote edges, achieving high throughput without full graph reconstruction.

    \item \textbf{Granularity-Adaptive Query Processing:} To mitigate the severe warp divergence caused by branch-heavy CPU filtering logic, we develop a filter-aware SIMT execution kernel. It combines early scalar pre-checks with a hybrid traversal strategy: utilizing remote edges for coarse-grained jumps over invalid ranges and local edges for fine-grained convergence. This ensures highly parallel, uniform thread execution across varying filter selectivities.
\end{itemize}
% \textbf{1) Bucket-Organized Hybrid Topology:} We partition data into ordered buckets based on scalar attributes and introduce a dual-layer edge structure (Local/Remote) to ensure graph connectivity under arbitrary filter conditions.
% \textbf{2) Two-Phase Construction Strategy:} We design a step-wise construction process that first establishes intra-bucket local connect and subsequently injects inter bucket remote links, effectively balancing local search precision with global navigability.
% \textbf{3) Dynamic Parallel Insertion:} We propose a batch-oriented insertion algorithm that leverages GPU massive parallelism to support simultaneous updates of both local and remote connections, achieving high insertion throughput without the need for full graph reconstruction.

% ==========================================
% 4.2 The Hardware-Aware Index Structure (Static View)
% 目标：只讲“是什么” (Definition, Properties, Layout)，不讲“怎么做”
% ==========================================
\subsection{Hardware-Aware Graph Topology}
\label{subsec:structure}

\subsubsection{Predicate-based Bucket Partitioning}
In conventional hybrid search, node-by-node predicate evaluation introduces irregular control flows that severely degrade SIMT execution efficiency. Concurrently, the random pointer-chasing inherent in graph traversal prevents coalesced memory reads, drastically underutilizing memory bandwidth. To resolve these hardware-algorithm mismatches, we shift the filtering paradigm: we replace fine-grained conditional branching with coarse-grained bucket selection, and physically store intra-bucket data contiguously to guarantee coalesced memory accesses during local traversal. Accordingly, given a dataset $\mathcal{D}$, we partition the objects into $m$ disjoint buckets $\mathcal{B} = \{B_1, B_2, \dots, B_m\}$ based on their scalar predicate $s_i$. This physical layout establishes a two-level hierarchy that optimizes both local execution and global navigation:

\begin{enumerate}
    \item \textbf{Intra-Bucket Efficiency:} 
    Within each bucket $B_i$, we construct a local proximity graph to accelerate search. The bucket capacity $N_{B_i}$ is carefully tuned to a ``sweet spot'' (e.g., $\sim 10^4$ vectors) where graph-based traversal $O(\log N_{B_i})$ significantly outperforms brute-force linear scanning $O(N_{B_i})$. More importantly, storing the vectors of a bucket continuously in memory ensures that graph traversal steps trigger coalesced memory accesses. Since adjacent threads in a warp likely access neighbors within the same dense bucket, this contiguous physical layout maximizes L2 cache hits and effective bandwidth utilization compared to random access patterns.

    \item \textbf{Inter-Bucket Management:} 
    From a global perspective, bucketing transforms range filtering into a coarse-grained selection problem. Given a query range $[l, u]$, we can identify the intersecting buckets in $O(1)$ time via a lightweight metadata map. This effectively substitutes complex GPU conditional branching with a simple mathematical range intersection. While this enables rapid pruning of irrelevant data, strict partitioning inherently fragments the graph. To address this, our partitioning scheme serves as the physical foundation for the Hybrid Edge Topology (Sec.~\ref{subsubsec:hybrid_edge}), where we explicitly overlay remote connections to restore navigability between these logically separated buckets.
\end{enumerate}

\subsubsection{Hybrid Edge Topology}
\label{subsubsec:hybrid_edge}
\begin{figure}[t]
    \centering
    \includegraphics[width=\linewidth]{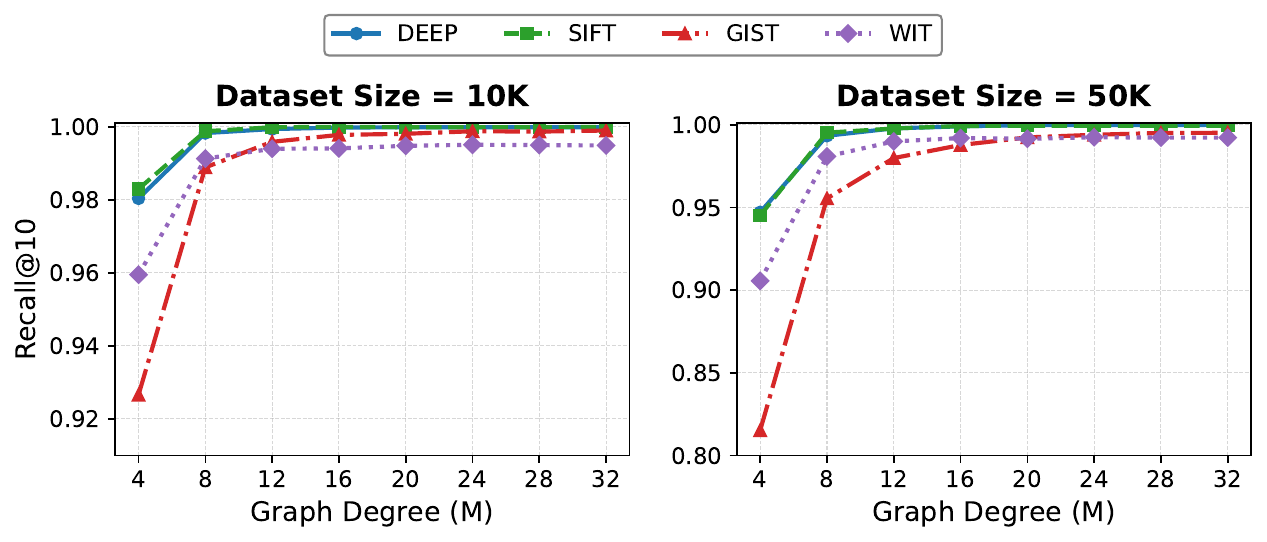}
    \caption{Impact of Graph Degree on Recall@10 for 10K and 50K datasets. In all cases, the search precision plateaus after a certain degree, demonstrating that increasing local connectivity yields marginal gains. }
    \label{fig:edgerecall}
\end{figure}

While partitioning maximizes locality, naively applying range filters fragments the graph, leading to the \emph{Connectivity Crisis}. 
To resolve this efficiently, we leverage a key empirical insight regarding \emph{Edge Redundancy}: within a moderate-sized bucket, the recall gains from increasing the local node degree exhibit diminishing returns, see Figure~\ref{fig:edgerecall}. 
This implies that a portion of the hardware-aware degree budget can be strategically repurposed to establish inter-bucket links without compromising local search precision.

Formally, we define the edge set $E(v)$ of a node $v \in B_i$ as the union of two functional subsets: $E(v) = E_{local}(v) \cup E_{remote}(v)$.
\begin{itemize}[leftmargin=*]
    \item \textbf{Local Edges ($E_{local}$):} These edges connect $v$ to its nearest neighbors \textit{within the same bucket} $B_i$. They preserve the local density of the high-dimensional manifold, ensuring high recall for fine-grained search within valid ranges.
    \item \textbf{Remote Edges ($E_{remote}$):} These edges connect $v$ to representative nodes in \textit{different buckets} $B_j (j \neq i)$. Mathematically, they function as "long-range links" in Small-World networks.
\end{itemize}
Crucially, remote edges act as \textit{bridges} that span across scalar intervals. They ensure that the greedy search (discussed in Sec.~\ref{subsec:query}) can \textit{jump} over buckets that might be filtered out, preserving global navigability under arbitrary query ranges.

\subsubsection{Hardware-Aware Degree Constraint}

To efficiently materialize this hybrid topology on GPUs, we must eliminate the architectural bottlenecks inherent in variable-degree graphs. Under the SIMT execution model, processing variable-length neighbor lists induces severe workload imbalance and intra-warp serialization. Furthermore, representing such irregular structures dictates indirect memory addressing (e.g., via offset arrays), which fundamentally disrupts coalesced memory accesses and throttles bandwidth utilization.

To eliminate these hardware inefficiencies, we impose a strict \emph{Fixed-Degree Constraint}: every node $u$ is allocated an identical total degree budget $K_{max}$ (e.g., $K_{max}=32$). Consequently, graph construction is reformulated as a strict budget allocation problem, governed by $|E_{local}| + |E_{remote}| = K_{max}$. This architectural constraint allows the entire graph's adjacency structure to be physically flattened into a dense, fixed-size 2D matrix. As a result, when a warp fetches neighbor lists, every thread executes the exact same number of memory reads from perfectly predictable, contiguous offsets. This structure guarantees alignment with GPU cache boundaries and ensures perfectly balanced thread execution, thereby maximizing throughput.

\subsubsection{Unified Memory Layout}
\label{subsubsec:memory_layout}

To manage this hardware-aligned graph efficiently while supporting high-throughput dynamic updates, \ourmethod adopts a \emph{Unified Global Layout}. Instead of maintaining fragmented, bucket-specific memory allocations, all data resides in two monolithic, row-major arrays allocated in the GPU device memory:
\begin{itemize}[leftmargin=*]
    \item Global Feature Matrix $\mathbf{X} \in \mathbb{R}^{N_{cap} \times d}$: Stores the high-dimensional base vectors, where $N_{cap}$ denotes the maximum allocated capacity.
    \item Global Adjacency Matrix $\mathbf{A} \in \mathbb{Z}^{N_{cap} \times K_{max}}$: Stores the fixed-degree neighbor lists, encompassing both local and remote edges.
\end{itemize}

Crucially, this monolithic design decouples logical bucket partitioning from physical storage. To seamlessly map physical row indices to logical buckets without physically re-sorting the massive global arrays during online data ingestion, we maintain two lightweight metadata tables: an Index-to-Bucket Map ($M_{I2B}$) that records the bucket ID for each physical node index, and a Bucket-to-Index Map ($M_{B2I}$) that tracks the set of physical node indices belonging to each logical bucket. Both maps share an $O(N)$ space complexity and easily reside in the GPU memory. This decoupled layout serves as the physical foundation that allows \ourmethod to perform zero-copy, append-only dynamic insertions (as detailed in Section~\ref{subsec:update}).

% \subsubsection{Memory Layout}
% % 简述数据在显存里怎么排布的 (Bucket-based storage)
% \ourmethod adopts a \textbf{Unified Global Layout} to manage graph data efficiently. All data resides in two monolithic arrays allocated in the GPU memory: Global Feature Matrix $\mathbf{X} \in \mathbb{R}^{N_{cap} \times d}$, stores the high dimensional vectors, Global Adjacency Matrix $\mathbf{A} \in \mathbb{Z}^{N_{cap} \times K}$, stores the fixed-degree neighbor lists.
% To manage the mapping between physical indices and logical buckets, we maintain two lightweight metadata tables: \textbf{Index-to-Bucket Map}, recording the bucket ID for each node, \textbf{Bucket-to-Index Map}, tracking the set of node indices belonging to each bucket. Their space complexity are all $O(N)$
% TODO: 描述 [Node Features | Adjacency List] 的连续存储方式

% ==========================================
% 4.3 Index Construction & Optimization (Dynamic View 1)
% 目标：只讲“怎么做” (Algorithm, Process, Formulas)
% ==========================================
\subsection{Two-Pass Index Construction}
\label{subsec:construction}

To efficiently construct the topology defined in Sec.~\ref{subsec:structure}, \ourmethod employs a \emph{Two-Pass Construction Strategy}, as illustrated in Figure~\ref{fig:construction}.

\subsubsection{Local Router Construction}
\label{subsubsec:phase1}

The first phase aims to establish high-density connections within each bucket to ensure accurate local search. To achieve this efficiently, we design a highly parallelized pipeline comprising exact $k$-NN initialization, batched iterative pruning, and a prioritized edge-merging strategy.

Leveraging our bucketing strategy where each partition typically contains fewer than 20,000 nodes, we optimize the initialization step by replacing the standard IVF-PQ\cite{10597683} method with a Brute-Force $k$-NN initialization. At this scale, brute-force computation on GPUs significantly outperforms approximate methods in speed while guaranteeing 100\% initial recall.

Furthermore, we maximize construction parallelism by batching the subsequent iterative pruning steps. Since local edges are strictly confined within bucket boundaries, the operations of reverse graph construction and edge merging are independent across buckets. This allows us to execute these steps using global kernels over the entire dataset, treating it as a batch of disjoint graphs rather than launching a fragmented kernel for each bucket.

In the final step of merging the forward and reverse graphs, we employ an interleaved merging strategy. This technique alternates between the two source lists to populate the final adjacency list, ensuring that the high-quality neighbors from both forward and reverse views are consolidated at the front. Consequently, the neighbors occupying the leading positions (Top-$K_{local}$) are designated as necessary edges, forming the backbone of local connectivity.

The remaining neighbors at the tail are classified as redundant edges. Instead, they serve as a fallback buffer: they are targeted for overwriting by remote edges in the subsequent phase. However, if the generated remote edges are insufficient to fill these slots, the original local redundant edges are preserved, providing supplementary connectivity to further robustify the graph.

\subsubsection{Remote Edge Construction}
\label{subsubsec:phase2}

While Sec.~\ref{subsubsec:phase1} guarantees dense intra-bucket connectivity, strict physical partitioning inherently leaves the buckets isolated, creating a macroscopic connectivity crisis under arbitrary range filters. The objective of this phase is to inject cross-bucket navigability without violating the rigid fixed-degree memory constraints required by the GPU.

The objective of this phase is to resolve the inter-bucket unreachability caused by partitioning. We construct a global graph to serve as a candidate pool for cross-bucket connections and fuse it with the local graph from Sec.~\ref{subsubsec:phase1}. Unlike Sec.\ref{subsubsec:phase1} which has relatively small data sizes, applying exact brute-force computation globally is computationally intractable ($O(N^2)$). Thus, we apply the standard CAGRA construction pipeline, in Alg.~\ref{alg:cagra_pipeline}, on the entire dataset $\mathcal{D}$ to handle the global scale efficiently. This involves IVF-PQ based initialization, iterative pruning, and reverse graph merging, resulting in a fully connected global graph $G_{global}$.

\begin{algorithm}[t]
\caption{Global Edge Construction Pipeline}
\label{alg:cagra_pipeline}
\begin{algorithmic}[1]
\REQUIRE Dataset $\mathcal{D}$, Degree Budget $K$
\ENSURE Optimized Graph Topology $G$

\STATE \textbf{// Step 1: Fast Initialization via IVF-PQ}
\STATE Train IVF-PQ codebooks on a sample of $\mathcal{D}$
\STATE $G_{init} \leftarrow \text{ApproximateKNN}(\mathcal{D}, \text{initial-}K)$ 

\STATE \textbf{// Step 2: Parallel Refinement \& Merging}
\STATE $G_{fwd} \leftarrow \text{NNDescentRefine}(G_{init})$ \COMMENT{Iterative optimization}
\STATE $G_{rev} \leftarrow \text{ConstructReverseGraph}(G_{fwd})$ \COMMENT{Enhance reachability}
\STATE $G_{merged} \leftarrow G_{fwd} \cup G_{rev}$

\STATE \textbf{// Step 3: Rank-based Selection (Fixed Degree)}
\FORALL{node $u \in G_{merged}$ \textbf{in parallel}}
    \STATE \textbf{// Keep only the closest $K$ neighbors to fit CSR layout}
    \STATE $N(u) \leftarrow \text{TopKByDistance}(u, G_{merged}[u], K)$
\ENDFOR

\RETURN $G \leftarrow \{N(u) \mid u \in \mathcal{D}\}$
\end{algorithmic}
\end{algorithm}

When selecting edges from $G_{global}$ to fill the redundant slots of a node $u$, we employ a strict filtering and ranking strategy. A naive global selection might populate all remote slots with neighbors from vastly distant buckets, making adjacent or nearby buckets unreachable during search. To address this, we partition the remote edge budget into two categories: (1) \emph{Proximal Remote Edges}, candidates whose scalar predicates fall within a neighboring range (e.g., $\pm 20\%$ of the total range). These ensure smooth transitions to adjacent buckets. (2) \emph{Global Remote Edges}, candidates sampled from the entire remaining scope to serve as long-range expressways. 

Finally, we perform the fusion. The selected remote edges are used to overwrite the Redundant Slots marked in Phase 1. Crucially, if the number of qualified remote edges is insufficient to fill the slots, we retain the original local redundant edges instead of leaving the slots empty. This fallback mechanism ensures that every edge slot contributes to graph connectivity, whether local or global.

% ==========================================
% 4.4 Arbitrary-Order Insertion (Dynamic View 2)
% 目标：讲动态更新逻辑，重点在于如何维持 32 度的约束
% ==========================================
\subsection{Dynamic Updates}
\label{subsec:insertion}

Support for high-throughput data ingestion is a critical requirement for modern vector databases, particularly for freshness-sensitive AI workloads like agentic memory and real-time recommendation. However, enabling dynamic updates within predicate-optimized graph indices presents a fundamental structural conflict. 

To accelerate predicate filtering, many existing designs tightly couple the data's physical storage layout with its scalar properties (e.g., maintaining physically sorted arrays)~\cite{zuo2024serf, wang2025timestamp}. Consequently, these memory layouts are inherently static. Inserting a new vector into such a tightly coupled structure inevitably forces the underlying data to shift. Because graph edges are fundamentally stored as physical memory offsets, any movement of the underlying data instantly invalidates all adjacency relationships pointing to the shifted nodes. We term this catastrophic failure the \emph{Cascading Index Shift}. Correcting these corrupted pointers requires a global scan and update of the entire adjacency matrix, incurring a computational overhead equivalent to a full graph reconstruction. This fundamental bottleneck makes prior physically-sorted layouts strictly incompatible with online ingestion.

\begin{figure}
    \centering
    \includegraphics[width=1.0\linewidth]{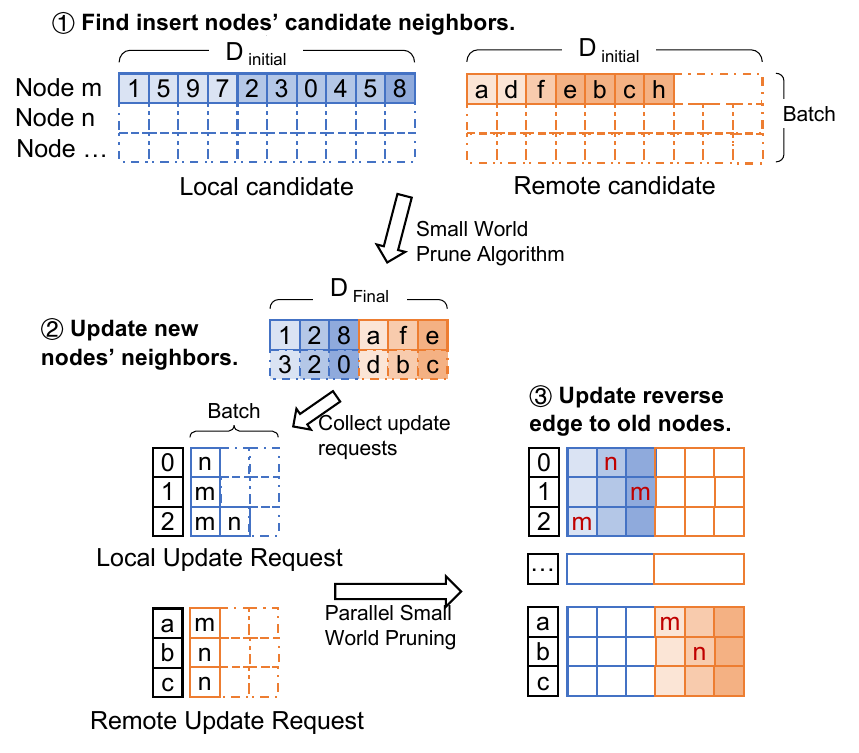}
    \caption{Update flow of \ourmethod for parallel insert.}
    \label{fig:insert}
\end{figure}

\subsubsection{Out-of-Order Insertion}

To resolve the cascading index shift and fully exploit GPU concurrency, \ourmethod leverages its decoupled \emph{Logical Bucketing Architecture} to execute Zero-Shift Append-Only Insertions. In the GPU memory hierarchy, dynamic mid-array reallocations or massive sequential shifts (e.g., \texttt{memmove} in VRAM) are prohibitively expensive operations that serialize execution and stall memory buses. When a new batch of data $\mathcal{Q}_{new}$ arrives, regardless of their scalar predicate values, we exploit massive SIMT parallelism to ingest them in $O(1)$ time:
\begin{enumerate}[leftmargin=*]
    \item Parallel Physical Append: Instead of acquiring complex write-locks, GPU threads utilize a single fast hardware atomic counter (e.g., \texttt{atomicAdd}) to concurrently claim continuous memory slots at the tail of the Global Feature and Adjacency Matrices. This lock-free, zero-shift operation guarantees that existing graph pointers remain strictly valid while maximizing VRAM write bandwidth.
    \item Asynchronous Logical Assignment: Concurrently, the threads update the lightweight metadata tables ($M_{I2B}$ and $M_{B2I}$) to map the newly acquired physical indices to their appropriate logical buckets. Because these metadata updates involve simple scalar writes without data movement, they execute seamlessly with minimal cache thrashing.
\end{enumerate}
This decoupled design ensures that \ourmethod maintains the search efficiency of predicate-aware bucketing without succumbing to the structural rigidity and parallel bottlenecks of physically sorted layouts.

% \ourmethod eliminates this bottleneck through its \textbf{Logical Bucketing Architecture}. By decoupling the logical order (buckets) from physical storage (global arrays), our system supports \textbf{arbitrary-order insertions} with zero data shifting.
% When a new batch of data $\mathcal{Q}_{new}$ arrives, regardless of their scalar predicate values, we perform an \textbf{Append-Only Physical Update}:
% \begin{enumerate}
%     \item \textbf{Physical Persistence:} New vectors are simply appended to the tail of the Global Feature/Adjacency Matrices. This is an $O(1)$ operation that requires no memory reorganization.
%     \item \textbf{Logical Assignment:} We concurrently update the lightweight metadata table $M_{I2B}$ to map the new physical indices to their appropriate logical buckets.
% \end{enumerate}
% This design ensures that \ourmethod maintains the efficiency of range filtering (via buckets) without the rigidity of physical sorting.

\subsubsection{Parallel Batch-Oriented Workflow}
Following the physical append operation, the graph topology must be updated to integrate the new nodes. In a GPU environment, processing insertions sequentially incurs prohibitive kernel launch overheads and severely underutilizes the massive SIMT compute units. To maximize instruction throughput and amortize these costs, \ourmethod processes topological updates in highly parallelized batches through a three-stage workflow:

% Following the physical persistence, we update the graph topology. To mitigate the overhead of GPU kernel launches and maximize instruction throughput, we process updates in batches rather than sequentially. The workflow proceeds in three parallel steps:

\begin{enumerate}[leftmargin=*]
    \item Parallel Candidate Search: For all $q \in \mathcal{Q}_{new}$, we launch a global kernel to identify candidate neighbors from both the local target bucket and the global scope.
    \item Forward Edge Selection: For each $q$, we select its optimal neighbors using \textit{Heuristic Pruning} (detailed below) and write them to the tail of the adjacency matrix.
    \item Parallel Reverse Rewiring: This step ensures graph connectivity (i.e., making $q$ reachable from existing nodes). Since graph edges are directed, we must identify existing nodes $\{v\}$ that should accept $q$ as a neighbor. To maximize parallelism, we first utilize a global atomic counter to collect all potential reverse update requests into a buffer, and then launch a dedicated kernel to process these rewiring operations concurrently.
\end{enumerate}

\subsubsection{Hardware-Constrained Topology Maintenance}

While heuristic pruning is a well-known concept in CPU-based indices like HNSW~\cite{8594636}, its application in \ourmethod is fundamentally mandated by our GPU hardware architecture. As established in Sec.~\ref{subsec:structure}, to guarantee coalesced memory accesses and eliminate warp divergence, we impose a strict fixed-degree constraint (e.g., $K_{max}=32$). 

This strict hardware boundary introduces a severe eviction problem during dynamic updates: inserting a new valid neighbor into an already saturated adjacency array necessitates the eviction of an existing one. If we naively retain only the absolute closest neighbors to satisfy this strict capacity, the graph rapidly degenerates into isolated, highly clustered cliques, severely degrading global navigability.

To preserve search accuracy under tight GPU memory constraints, we strictly prioritize \emph{spatial diversity} over absolute proximity. For a source node $v$ and a new candidate $c$, $c$ is permitted to overwrite an existing neighbor slot $\mathcal{N}(v)$ if and only if it provides a topological pathway not already covered by existing neighbors. Formally, we apply the Small-World pruning condition:
\begin{equation}
    \label{eq:standard_pruning}
    \text{dist}(v, c) < \text{dist}(c, n), \quad \forall n \in \mathcal{N}(v)
\end{equation}
\ourmethod ensures that the limited, fixed-size adjacency slots enforced by the GPU are strictly reserved for diverse, long-range connections. This prevents topological collapse and maintains the navigability essential for high-recall routing, completely offsetting the structural compromises demanded by the hardware.

% \subsubsection{Topology-Preserving Edge Maintenance}
% A core challenge in fixed-degree graphs is the \textbf{Eviction Problem}: inserting a new neighbor into a full adjacency list requires removing an existing one. Simply keeping the $K$ closest candidates leads to clustering and loss of global navigability. 

% To mitigate this, we adopt the \textbf{Small World Pruning} strategy from HNSW~\cite{8594636}, which prioritizes spatial diversity. Specifically, for a source node $v$, a candidate $c$ is added to the neighbor set $\mathcal{N}(v)$ if and only if it is closer to $v$ than to any already selected neighbor $n \in \mathcal{N}(v)$. Formally, the condition is:
% \begin{equation}
%     \label{eq:standard_pruning}
%     \text{dist}(v, c) < \text{dist}(c, n), \quad \forall n \in \mathcal{N}(v)
% \end{equation}
% This ensures that the graph retains long-range edges essential for the "Small World" property.

\textit{Freshness-Biased Pruning.}
While effective for static construction, Eq.~\ref{eq:standard_pruning} becomes a barrier during dynamic updates. Existing neighbors often form a geometric ``shield'' around $v$, causing the condition to fail for newly inserted nodes even if they are geometrically close. We term this the "Isolation of Newcomers".

To counteract this, we propose a bias-aware modification. We substitute the physical distance on the left-hand side of Eq.~\ref{eq:standard_pruning} with an Effective Distance ($d_{eff}$):
\begin{equation}
    \label{eq:freshness_bias}
    d_{eff}(v, x) = 
    \begin{cases} 
    \alpha \cdot \text{dist}(v, x), & \text{if } x \in \mathcal{Q}_{new} \text{ (fresh node)} \\
    \text{dist}(v, x), & \text{otherwise}
    \end{cases}
\end{equation}
where $\alpha \in (0, 1)$ is a bias coefficient. 

By using $d_{eff}(v, c)$ in the diversity check, we artificially increase a new node's probability of satisfying the condition. This strategy forces the graph to "open up" for newcomers, ensuring they are rapidly integrated into high-speed navigation paths without requiring global reconstruction.

% ==========================================
% 4.5 Granularity-Adaptive Query Processing
% 目标：讲查询逻辑，如何利用 Remote Edge 跳跃
% ==========================================
\subsection{Granularity-Adaptive Query Processing}
\label{subsec:query}

\begin{algorithm}[t]
\caption{\ourmethod Range-Constrained Search}
\label{alg:search}
\begin{algorithmic}[1]
\REQUIRE Query $q$, Range $[l, u]$, Graph $\mathbf{A}$, Data $\mathbf{X}$, Metadata $M_{B2I}$
\ENSURE Top-$k$ results $\mathcal{R}$

\STATE $B_{valid} \leftarrow \text{GetIntersectingBuckets}(M_{B2I}, [l, u])$
\STATE $\mathcal{S} \leftarrow \text{SampleSeedsFrom}(B_{valid})$
\STATE $\mathcal{Q} \leftarrow \text{ComputeDist}(q, \mathcal{S})$

\WHILE{not converged}
    \STATE $P \leftarrow \text{PopUnvisited}(\mathcal{Q})$
    \STATE $\mathcal{N} \leftarrow \text{GatherNeighbors}(\mathbf{A}, P)$
    
    \FORALL{candidate $v \in \mathcal{N}$ \textbf{in parallel}}
        \IF{$s_v < l$ \textbf{or} $s_v > u$}
            \STATE \textbf{continue} 
        \ENDIF
        
        \STATE $d \leftarrow \text{ComputeDistance}(q, v)$
        \STATE $\text{UpdateQueue}(\mathcal{Q}, v, d)$
    \ENDFOR
\ENDWHILE

\STATE \RETURN Top-$k$ from $\mathcal{Q}$
\end{algorithmic}
\end{algorithm}

% Algorithm~\ref{alg:search} outlines the execution flow of our query processing. To achieve maximum throughput, we adopt a warp-centric execution model similar to CAGRA ~\cite{10597683}, but we introduce critical range-aware mechanisms that enable \textbf{Granularity-Adaptive Navigation}.

Algorithm~\ref{alg:search} outlines the execution flow of our query processing. To maximize throughput and guarantee high recall under arbitrary scalar constraints, we introduce two critical range-aware mechanisms:

Standard graph search typically initializes with random global seeds or a fixed entry point~\cite{malkov2014approximate,jayaram2019diskann}. In filtered scenarios, this is highly inefficient: seeds often fall into ``invalid range'' disjoint from the query range $[l, u]$, requiring long, fruitless traversal paths just to reach the valid manifold. \ourmethod eliminates this overhead by exploiting the lightweight metadata $M_{B2I}$ to perform an $O(1)$ \emph{Targeted Initialization} (Alg.~\ref{alg:search}, Lines 1-3). By sampling entry points strictly from physical buckets that intersect $[l, u]$, we ensure the search initiates immediately within the ``hot zone'' of the high-dimensional space.

During greedy expansion, we strictly enforce the range constraint via a \emph{Scalar Pre-check} (Alg.~\ref{alg:search}, Lines 9-11). Computing high-dimensional vector distances is exceptionally memory- and compute-intensive ($O(d)$ complexity). By contrast, evaluating a scalar predicate is fundamentally a cheap $O(1)$ register-level operation. Therefore, candidates falling outside $[l, u]$ are discarded immediately, completely bypassing the expensive vector math. 

In standard monolithic graphs, such aggressive pre-filtering would sever the search path and cause the routing to fall into dead ends. However, our hybrid topology guarantees safe navigation:
\begin{itemize}[leftmargin=*]
    \item \emph{Bridge Traversal (Coarse-grained):} Even if massive swaths of intermediate nodes are pruned by the filter, the remote edges ($E_{remote}$) act as bridges, allowing the search to ``jump'' over invalid regions and land safely on promising valid buckets.
    \item \emph{Local Refinement (Fine-grained):} Once inside a valid bucket, the dense local edges ($E_{local}$) enable the greedy search to rapidly converge to the exact nearest neighbors.
\end{itemize}

\section{Experiments}
\label{sec:experiments}

In this section, we conduct a comprehensive evaluation to demonstrate the efficiency, robustness, and scalability of \ourmethod. Our experiments are designed to answer the following key research questions:

\begin{itemize}[leftmargin=*]
    \item \textbf{RQ1. End-to-End Search Efficiency:} How does \ourmethod compare against state-of-the-art hybrid search baselines in terms of query throughput (QPS) across diverse predicate selectivities?
    
    \item \textbf{RQ2. Index Construction \& Dynamic Updates:} How efficient is our index construction, and can the insertion sustain high-throughput online ingestion without compromising search accuracy?
    
    \item \textbf{RQ3. Topological Integrity:} Can \ourmethod construction and insertion method maintain the connectivity of the graph?
    
    \item \textbf{RQ4. Scalability:} How does the system scale with increasing dataset size?
    
    \item \textbf{RQ5. Ablation:} What are the individual performance contributions of our core design choices?
\end{itemize}

% In this section, we conduct a comprehensive evaluation to demonstrate the efficiency, robustness, and scalability of \ourmethod. Our experiments aim to answer the following key research questions:
% \begin{itemize}
%     \item \textbf{RQ1.Search Efficiency:} How does \ourmethod compare against state-of-the-art baselines in terms of Query Per Second (QPS) under various filtering selectivity?
%     \item \textbf{RQ2.Update Performance:} Can \ourmethod support high-concurrency dynamic updates (build time and insert time for indexing) without compressing search quality?
%     \item \textbf{RQ3.Topological Integrity Analysis:} Can \ourmethod construction and insertion method maintain the connectivity of the graph?
%     \item \textbf{RQ4.Scalability:} How does the system scale with increasing dataset size?
%     \item \textbf{RQ5.Ablation:} What is the contribution of the Hybrid Topology?
% \end{itemize}

\subsection{Setup}

All experiments were conducted on a workstation equipped with an \textbf{Intel Core i9-14900 CPU}, \textbf{128 GB DDR5 RAM}, and an \textbf{NVIDIA GeForce 3090Ti GPU} (24 GB VRAM), running Ubuntu 24.04 LTS.
\ourmethod and GPU baselines are implemented in C++ with CUDA 12.0.
All CPU baselines are implemented in C++ and parallelized with OpenMP.
To ensure fairness across methods, we use a fixed configuration of \textbf{32 threads} for both index construction and query processing unless otherwise specified.
For throughput (QPS), we issue concurrent queries equal to the number of query threads and report the aggregate QPS; for build time, we report wall-clock time under the same parallel setting.

\subsubsection{Datasets}
\label{sec:dataset}
We evaluate our approach on four standard high-dimensional vector benchmarks varying in scale and dimensionality. The statistics are summarized in Table~\ref{tab:datasets}.
Since these public datasets contain only vector data, we follow the established protocol in recent studies to generate synthetic scalar predicates. Specifically, we assign a random scalar value $s \in [0, 1]$ (simulating timestamps or prices) to each vector following a uniform distribution.

\begin{itemize}[leftmargin=*]
    \item \textbf{DEEP-96~\cite{johnson2019billion}}: 
    Deep visual feature vectors extracted from a GoogLeNet model trained on ImageNet, following the standard DEEP benchmark used in large-scale similarity search.

    \item \textbf{SIFT-128~\cite{jegou2011product}}: 
    A classic benchmark dataset consisting of local SIFT image descriptors, widely used to represent low-dimensional vector search workloads.

    \item \textbf{GIST-960~\cite{jegou2011product}}: 
    A benchmark dataset of global GIST image descriptors that capture holistic scene information, representing high-dimensional workloads.

    \item \textbf{WIT-2048~\cite{srinivasan2021wit}}: 
    High-dimensional image embeddings derived from the Wikipedia Image Text (WIT) dataset, representing ultra-high-dimensional vector workloads.
\end{itemize}

\begin{figure*}
    \centering
    \includegraphics[width=1.0\linewidth]{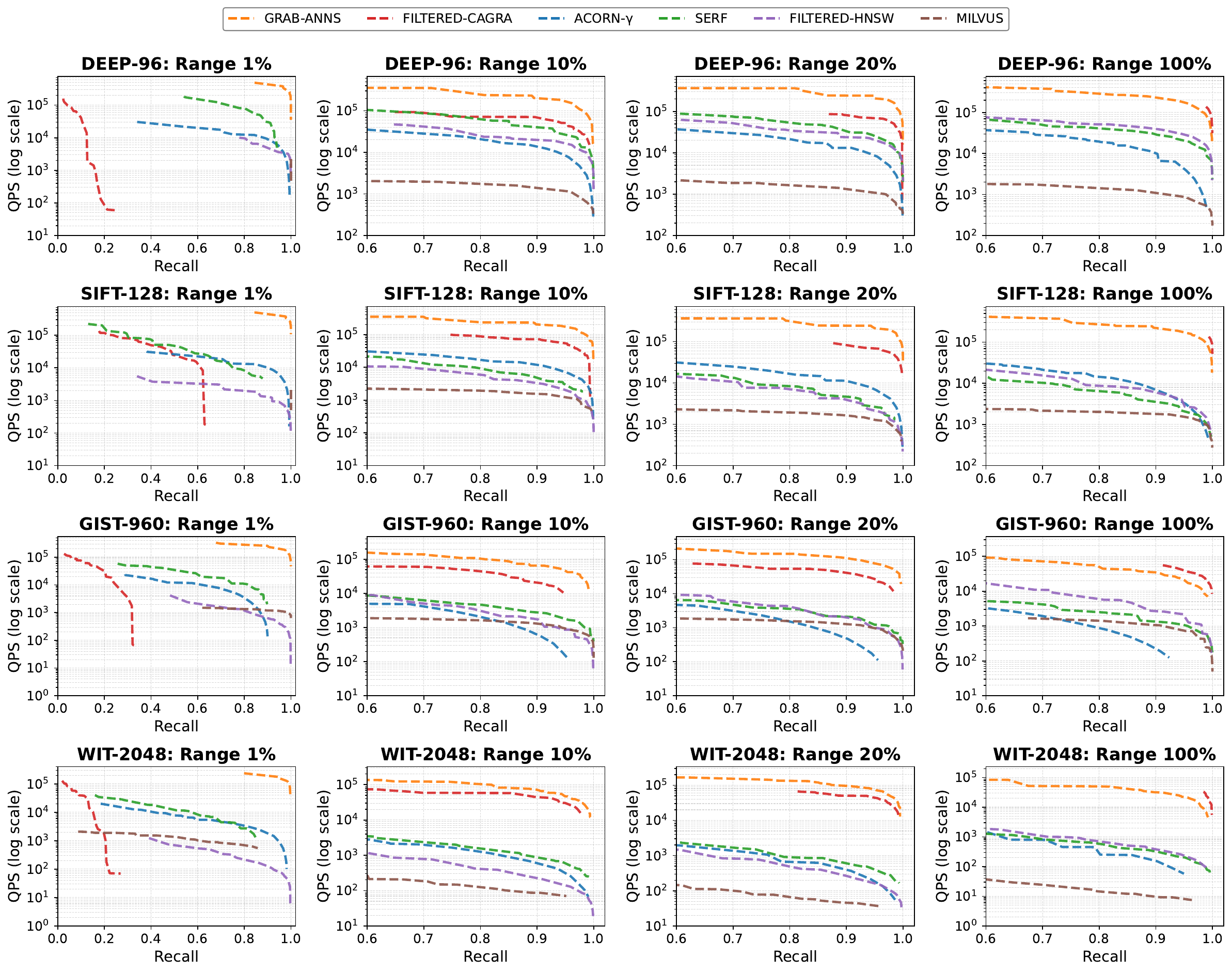}
    \caption{The performance of \ourmethod and baselines, testing in four datasets and four ranges}
    \label{fig:compare2baseline}
\end{figure*}

\begin{table}[t]
    \centering
    \caption{Dataset Statistics}
    \label{tab:datasets}
    \begin{tabular}{l|c|c|c}
    \toprule
    \textbf{Dataset} & \textbf{Dimensions} & \textbf{Base Size} & \textbf{Metric} \\
    \midrule
    DEEP-96      & 96   & 1,000,000 & Euclidean \\
    SIFT-128      & 128  & 1,000,000 & Euclidean \\
    GIST-960      & 960  & 1,000,000 & Euclidean \\
    WIT-2048   & 2048 & 1,000,000 & Euclidean \\
    \bottomrule
    \end{tabular}
\end{table}

\subsubsection{Baselines}
We compare \ourmethod against four representative baselines covering both algorithmic and system-level state-of-the-arts:
\begin{itemize}[leftmargin=*]
    \item \textbf{HNSW-Filter~\cite{wang2025timestamp}:}
    An ANNS-first baseline based on HNSW with a pre-filtering strategy,
    where range constraints are integrated into the search procedure.
    We use the optimized implementation from HNSWlib, which implements
    the HNSW algorithm~\cite{malkov2018efficient}.

    \item \textbf{SeRF~\cite{zuo2024serf}:} A state-of-the-art filtered graph index for efficient range queries. We adopt its 2D Segment Graph variant with the MaxLeap (rightmost-neighbor) strategy, as discussed in the original paper.
    
    \item \textbf{Milvus~\cite{wang2021milvus}:} A widely adopted open-source vector database. We deploy Milvus in standalone mode using its HNSW index engine, and perform range filtering via scalar predicate filtering (boolean expressions) on the primary-key field (\texttt{id}) during search.

    \item \textbf{ACORN~\cite{10.1145/3654923}:}
    A predicate-agnostic filtered ANN index built on FAISS.
    It builds an ACORN graph index and performs filtered search by restricting graph traversal to points satisfying the query predicate.
    Following the original paper, we set $\gamma = 100$ according to the minimum selectivity of 1\%, and use $M_\beta = 2M$ during index construction. We evaluate ACORN-$\gamma$ using the reference implementation.

    \item \textbf{CAGRA-Filter:}
    An adapted baseline built upon CAGRA~\cite{10597683}, the SOTA GPU-accelerated graph index. To evaluate its performance under range constraints, we explicitly augment its warp-centric search kernel with a pre-filtering mechanism. 
\end{itemize}

\subsubsection{Evaluation Protocol}
For all methods, we measure performance under a Top-10 recall target (Recall@10).
To strictly evaluate throughput, search queries are processed in batches of 100 for all approaches.
The filtering range $[l, u]$ is randomly generated with varying selectivities (e.g., covering 1\%, 10\%, 20\% 100\% of the dataset) to test robustness across different query scenarios.

To obtain the Recall--QPS Pareto frontiers, we evaluate each method under multiple index construction and search parameter configurations.
For graph-based CPU baselines, we vary the graph connectivity parameter $M$ and the construction parameter $\texttt{ef\_con}$, and sweep $\texttt{ef\_search}$ at query time to explore different accuracy--throughput trade-offs.
Specifically, we test $M \in \{8, 16, 32, 64\}$ with $\texttt{ef\_con} \in \{100, 200, 400, 800\}$, and $\texttt{ef\_search}$ ranging from 16 to 2048.
All reported Pareto frontiers are constructed from the union of these configurations.

%%%%%%%%%%%%%%%%%%%
\subsection{RQ1. Search Efficiency}
Figure~\ref{fig:compare2baseline} presents the Recall--QPS Pareto frontiers on four datasets (DEEP-96, SIFT-128, GIST-960, WIT-2048) under four selectivities (1\%, 10\%, 20\%, 100\%).
The QPS axis uses a logarithmic scale due to large performance gaps.
At Recall@95\%, we report the maximum QPS among configurations with Recall@10 $\ge 0.95$.

\subsubsection{Overall Performance Against CPU Baselines}
We compare against CPU baselines, including algorithmic baselines (SeRF, HNSW-Filter, ACORN-$\gamma$) and a system-level baseline (Milvus).
Overall, \ourmethod lies on the Pareto frontier and consistently dominates these CPU baselines at comparable recall.
Across 12 (dataset, selectivity) settings for 10\%/20\%/100\% selectivities where at least one CPU baseline reaches Recall@95\%, the speedup over the strongest CPU baseline (best of SeRF/HNSW-Filter/ACORN-$\gamma$/Milvus per setting) ranges from \textbf{6.46$\times$ to 240.10$\times$}, with a median of \textbf{28.16$\times$}.

For example, on DEEP-96 with 10\% selectivity, \ourmethod achieves 152{,}535 QPS at Recall@10=0.9613, compared to 23{,}174 QPS for SeRF (0.9690) and 1{,}136 QPS for Milvus (0.9596), corresponding to 6.58$\times$ and 134.28$\times$ speedups, respectively.

\subsubsection{Comparison with the GPU Baseline (Filtered-CAGRA)}
We further compare against the GPU baseline Filtered-CAGRA under selective search (range=10\% and 20\%).
Under these settings, \ourmethod consistently outperforms Filtered-CAGRA at Recall@95\%.
Across 6 comparable (dataset, selectivity) settings, the speedup ranges from \textbf{1.29$\times$ to 9.91$\times$}, with a median of \textbf{4.21$\times$}.

For instance, on SIFT-128 with 20\% selectivity, \ourmethod reaches 210{,}066 QPS (Recall@10=0.9504), while Filtered-CAGRA reaches 21{,}206 QPS (Recall@10=0.9594), giving a \textbf{9.91$\times$} speedup.

\subsubsection{Robustness Across Selectivities}
Across datasets and selective search regimes, \ourmethod maintains high throughput at Recall@95\% and remains consistently faster than CPU baselines.
On WIT-2048 with 20\% selectivity, \ourmethod reaches 73{,}712 QPS at Recall@10=0.9517, while SeRF reaches 307 QPS (0.9673), yielding a \textbf{240.10$\times$} speedup.
Compared with Filtered-CAGRA, \ourmethod also preserves clear gains under selective settings, including 1.29$\times$ on WIT-2048 (10\%) and 5.30$\times$ on WIT-2048 (20\%).

\subsubsection{Impact of Dimensionality}
The advantage over CPU baselines becomes more pronounced in high dimensions.
Under 10\% selectivity at Recall@95\%, SeRF drops from 23{,}174 QPS on DEEP-96 to 427 QPS on WIT-2048 (1.84\% retained), and HNSW-Filter drops from 15{,}090 QPS to 129 QPS (0.85\% retained).
Milvus drops from 1{,}136 QPS to 70 QPS (6.16\% retained), and ACORN-$\gamma$ drops from 7{,}699 QPS to 256 QPS (3.33\% retained).
In contrast, \ourmethod decreases from 152{,}535 QPS to 46{,}443 QPS (30.45\% retained), indicating that GPU parallelism better amortizes high-dimensional distance computation and memory traffic.

%%%%%%%%%%%%%%%%%%%%%%%%%%
\subsection{RQ2.Update Performance}
\label{subsec:update}

To evaluate the dynamic capabilities of \ourmethod, we analyze two critical aspects: \textit{Insertion Throughput} (how fast data can be ingested) and \textit{Index Stability} (how search quality evolves during continuous insertion).

\subsubsection{High-Throughput Ingestion}

\begin{figure}
    \centering
    \includegraphics[width=1.0\linewidth]{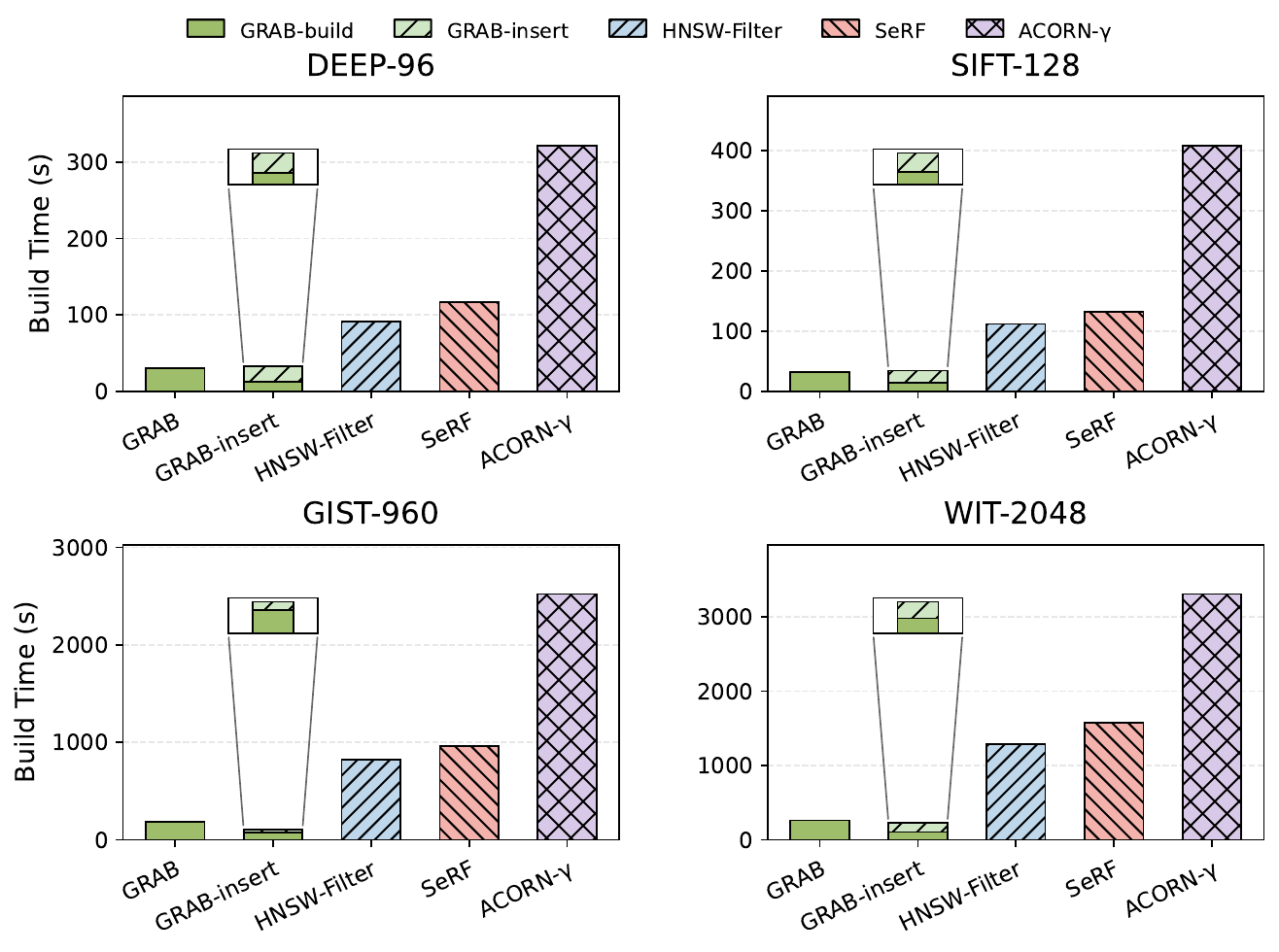}
    \caption{\textbf{Index Construction Comparison.} Total time to ingest 1M vectors. \textit{\ourmethod{}-insert} combines base indexing (blue) and dynamic insertion (red) to achieve bulk-build-class ingestion latency.}
    \label{fig:consturction_time}
\end{figure}

Figure~\ref{fig:consturction_time} compares the total time to ingest 1M vectors across four datasets. We define \textit{Build Time} as the wall-clock time required to reach 1M vectors in the index. For the CPU baselines (HNSW, SeRF, ACORN-$\gamma$) and \ourmethod, this corresponds to a static bulk build. For \ourmethod{}-insert, we adopt a hybrid protocol to rigorously test dynamic performance: the index is initialized with a base set (using \ourmethod bulk build), and the remaining data is ingested via our Parallel Batch Insertion mechanism.

As shown in Figure~\ref{fig:consturction_time}, \ourmethod{}-insert ingests 1M vectors in \textbf{32.7--226.2} seconds across datasets, yielding a speedup of \textbf{2.80$\times$ to 23.76$\times$} over the CPU baselines (HNSW, SeRF, ACORN-$\gamma$). For instance, on GIST-960, ingesting 1M vectors requires 959.4 seconds for SeRF, while \ourmethod{}-insert completes the task in 106.2 seconds (\textbf{9.03$\times$} faster). On the same dataset, ACORN-$\gamma$ requires 2523.1 seconds, making \ourmethod{}-insert \textbf{23.76$\times$} faster.

Remarkably, the total latency of \ourmethod{}-insert (Base Index + Dynamic Insert) remains comparable to the static bulk build of \ourmethod: 32.7 vs.\ 30.2 seconds on DEEP-96 (+8.2\%), and 226.2 vs.\ 262.8 seconds on WIT-2048 ($-13.9$\%). This finding is critical: it indicates that our \textbf{Append-Only Layout} and \textbf{Freshness-Biased Pruning} incur minimal overhead compared to highly optimized static CSR construction. Consequently, \ourmethod successfully brings the flexibility of dynamic updates to GPUs without the heavy performance penalty typically associated with dynamic data structures.

\subsubsection{Stability Under Continuous Updates}
To verify that our dynamic updates do not degrade the graph topology (e.g., creating "orphan" nodes), we conducted a stability test on the SFIT dataset.
The experiment begins with a pre-loaded index containing 50\% of the data (500k vectors). we then continuously insert the remaining 500k vectors in batches, measuring the Search QPS (at a fixed Recall $\ge$ 95\%) at every 50k interval.

\begin{figure}
    \centering
    \includegraphics[width=1.0\linewidth]{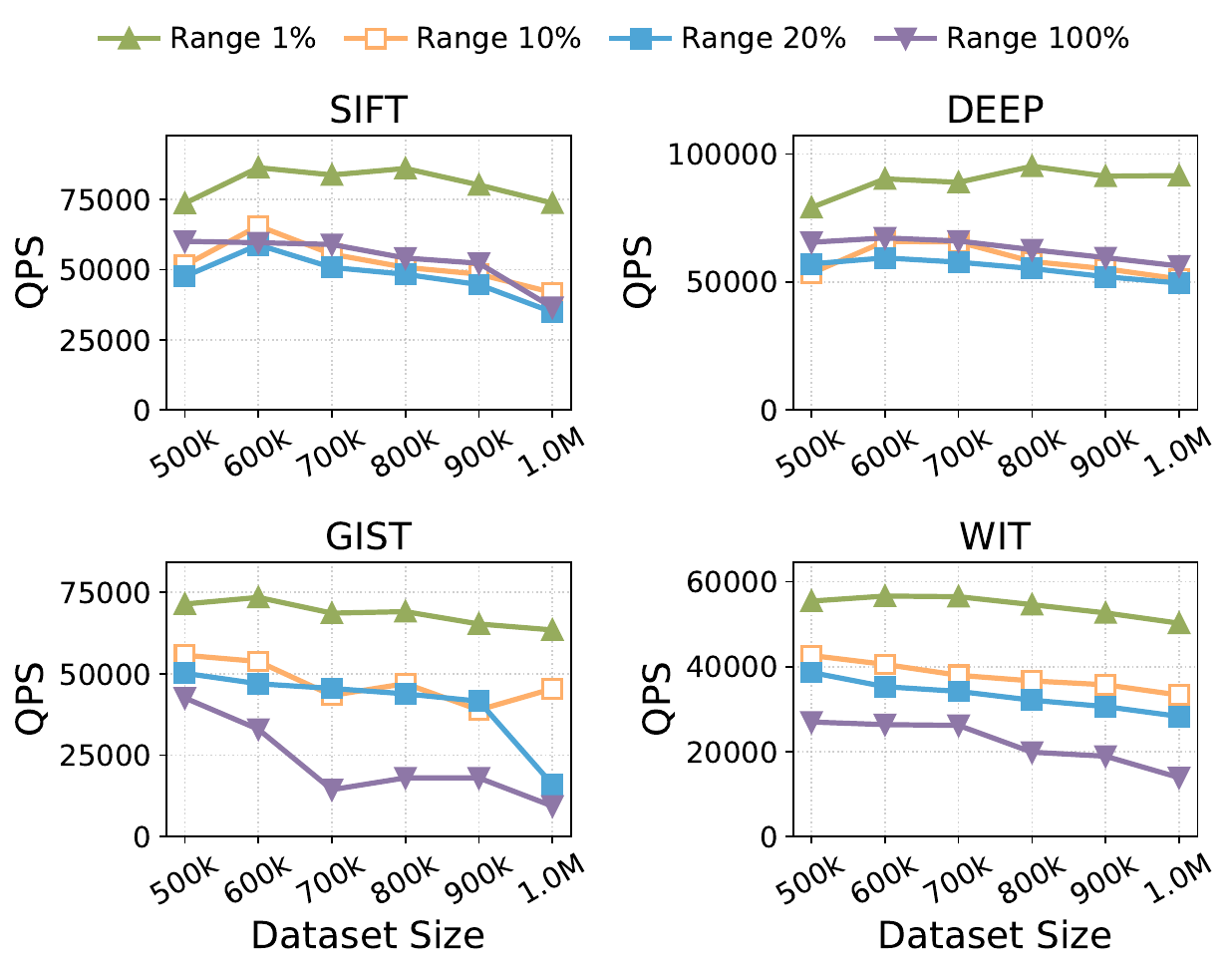}
    \caption{\textbf{Search Stability on SIFT-128.} QPS performance remains stable across varying range ratios (1\%-100\%) as the dataset size doubles from 0.5M to 1M via dynamic batch insertion.}
    \label{fig:stability}
\end{figure}

Figure~\ref{fig:stability} illustrates the performance trend during continuous ingestion. As the data size grows from 0.5M to 1M, the QPS exhibits a gentle, sub-linear decline. This behavior is theoretically expected, as the graph search complexity scales logarithmically ($O(\log N)$) with the dataset size $N$. 
Crucially, the absence of sudden performance drops confirms that our \textbf{Freshness-Biased Pruning} strategy effectively preserves the graph topology by successfully integrating new nodes into navigable paths. Even after doubling the dataset size via dynamic insertion, \ourmethod maintains consistently high throughput without degradation, demonstrating its suitability for long-running online services.

%%%%%%%%%%%%%%%%%%%%%%%%%%%%%%%%%
\subsection{RQ3.Topological Integrity Analysis}
\label{sec:q3_topological}

\begin{figure}
    \centering
    \includegraphics[width=1.0\linewidth]{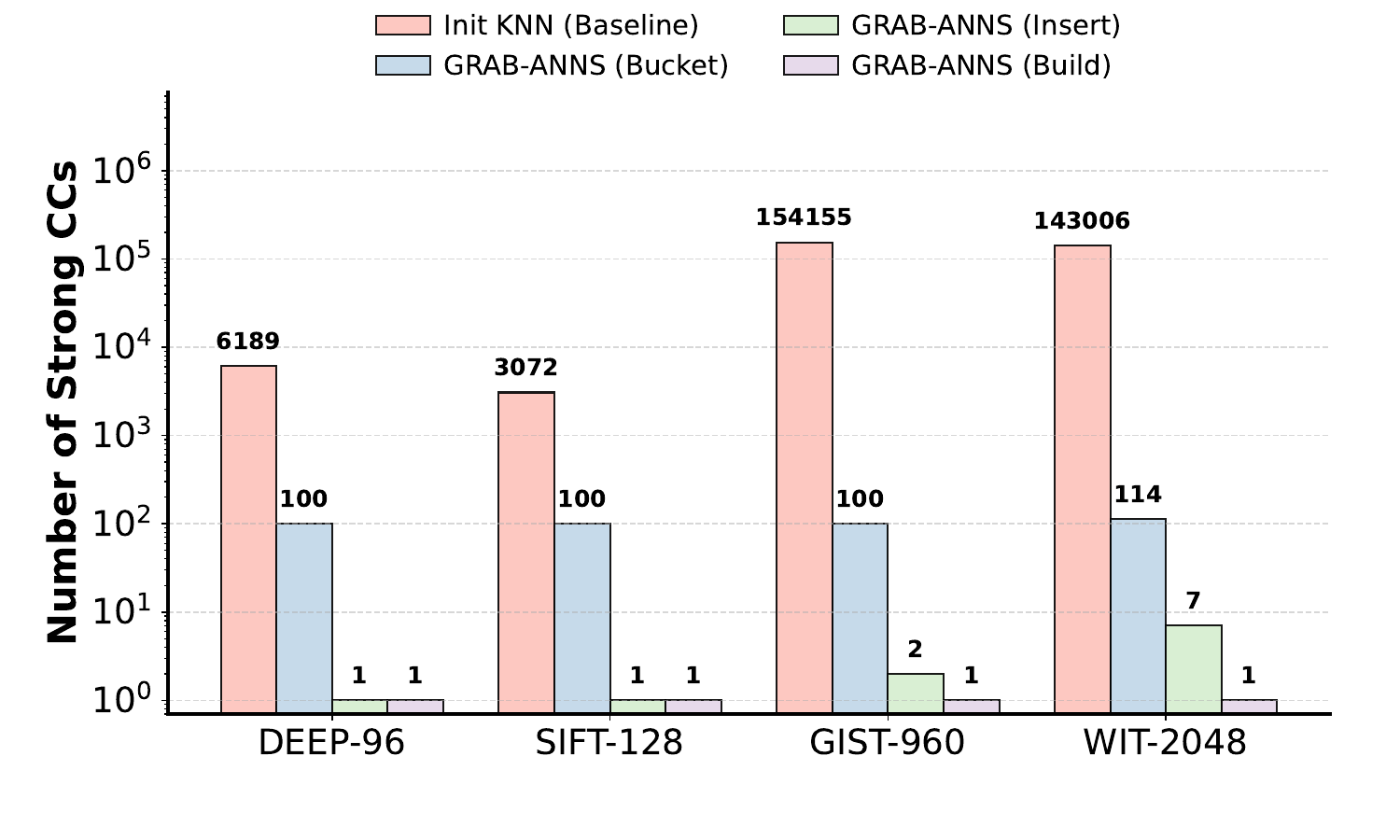}
    \caption{The behavior of \ourmethod build and update algorithm, using strongly connected components(scc) to show the graph quality. The fewer scc, the graph better, ideally 1. We compare on different datasets with different dimension and vector space distribution. For \ourmethod-bucket, we just skip the phase 2 mentioned in \ref{subsec:construction}. For \ourmethod-insert, we build base graph use 50\% data, and insert left 50\% data.}
    \label{fig:strong_cc}
\end{figure}

To rigorously verify that both our construction and update strategies maintain a high-quality graph structure, we analyze the \textbf{Number of Strongly Connected Components (Strong CCs)}. 
In a directed $k$-NN graph, topological integrity is quantified by navigability: an ideal graph consists of a single Strong CC, implying that any node is reachable from any other node. Conversely, a high number of CCs indicates fragmentation into "islands," causing the greedy search to terminate prematurely in local optima.

Figure~\ref{fig:strong_cc} dissects the connectivity across four scenarios:
\begin{enumerate}
    \item \textbf{Init KNN (Baseline):} A naive $k$-NN graph constructed without reverse edge reinforcement or global wiring.
    \item \textbf{\ourmethod (Bucket):} The graph after \textit{Phase 1} of construction only. It establishes dense local connections within buckets but lacks \ref{subsubsec:phase2} (remote edges).
    \item \textbf{\ourmethod (Insert):} The graph resulting from a hybrid workload: initializing with 50\% data and dynamically inserting the remaining 50\% using our Freshness-Biased strategy.
    \item \textbf{\ourmethod (Build):} The complete static graph constructed via the full Two-Phase process.
\end{enumerate}

\textbf{Analysis of Construction Robustness.}
The init $k$-NN baseline suffers from catastrophic fragmentation (e.g., $>10^5$ CCs on GIST), rendering it unusable for search. 
\ourmethod (Bucket) significantly improves connectivity but still leaves approximately 100 isolated components across datasets. This empirical evidence validates the "Connectivity Crisis" hypothesis: partitioning data into buckets inevitably severs global pathways.
However, once \ref{subsubsec:phase2} is applied (\ourmethod-Build), the graph achieves perfect or near-perfect connectivity (1 CC), proving that our remote edge wiring successfully bridges these partitioned islands.

\textbf{Analysis of Update Robustness.}
Crucially, \ourmethod (Insert) demonstrates exceptional stability. Despite ingesting 50\% of the data dynamically, the graph maintains a very low number of CCs (e.g., 20 on DEEP, 1 on GIST), which is orders of magnitude better than the naive baseline and significantly better than the Bucket-only graph. 
This confirms that our dynamic insertion mechanism effectively functions as an online version of "Remote Wiring",  proactively healing the topology and preventing new nodes from becoming isolated.

%%%%%%%%%%%%%%%%%%%%%%%%%%%%%%%%
\subsection{RQ4: Scalability}
\label{sec:q4_scalability}

\begin{figure}[t]
    \centering
    \includegraphics[width=1.0\linewidth]{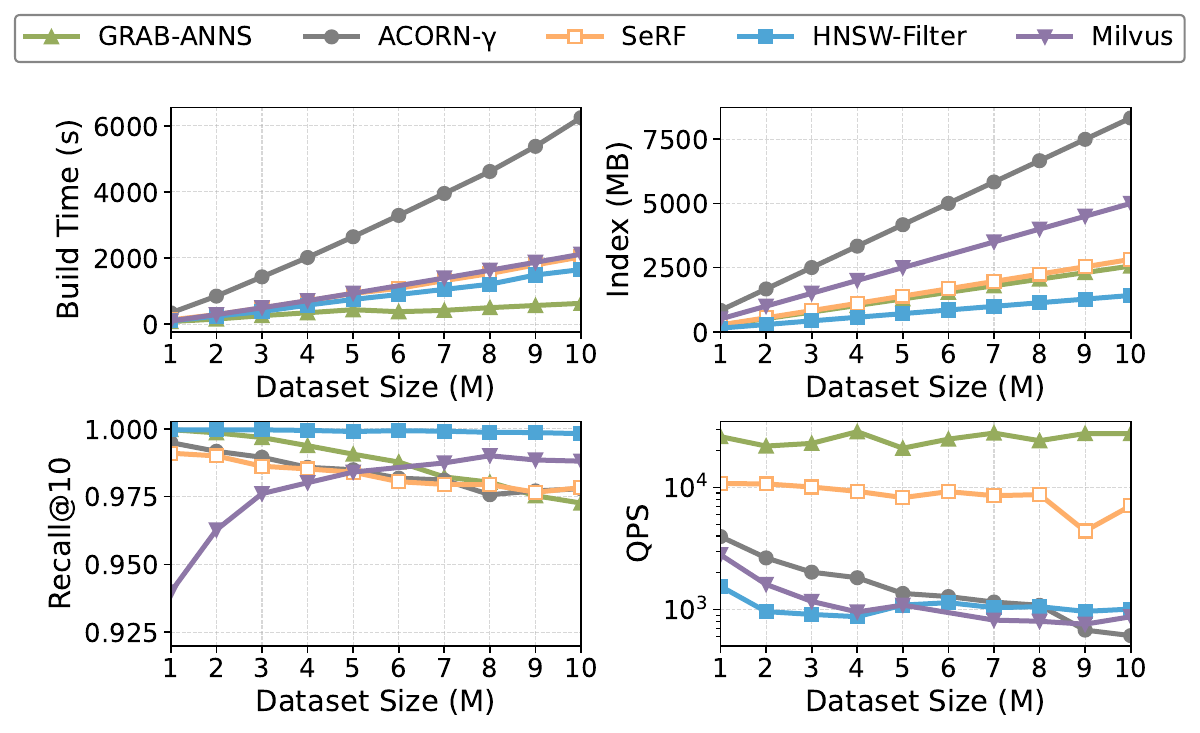}
    \caption{Scalability on DEEP-96 as the corpus size increases from 1M to 10M with \emph{fixed} hyperparameters. We report (top-left) build time, (top-right) graph-only index size (excluding base vectors), (bottom-left) Recall@10 at range $=10\%$, and (bottom-right) query throughput (QPS, log scale). \ourmethod uses $M{=}64$, $\textit{itop\_k}{=}512$, and $\textit{max\_iterations}{=}100$; ACORN-$\gamma$, filtered-HNSW, SeRF, and Milvus use $M{=}32$, $K{=}400$, and $K_{\text{Search}}{=}256$.}
    \label{fig:scal}
\end{figure}

We evaluate scalability on DEEP-96 by increasing the corpus size from 1M to 10M while keeping each method's hyperparameters fixed.
This isolates the effect of $N$ and tests whether performance can be preserved without per-scale retuning.
Figure~\ref{fig:scal} reports build time, graph-only memory footprint, Recall@10 at range $=10\%$, and QPS.

\subsubsection{Build-time scalability}
Figure~\ref{fig:scal} (top-left) shows that build time increases with $N$ for all methods, but \ourmethod exhibits the smallest growth rate.
From 1M to 10M, \ourmethod increases by $8.4\times$ (74.5\,s to 626.5\,s), whereas ACORN-$\gamma$ increases by $18.5\times$ (337.8\,s to 6245.5\,s).
Filtered-HNSW, SeRF, and Milvus increase by $17.4\times$ (94.9\,s to 1648.7\,s), $16.3\times$ (124.7\,s to 2032.1\,s), and $22.7\times$ (93.2\,s to 2112.5\,s), respectively.
At 10M, \ourmethod completes construction in 626.5\,s (10.4\,min), while the baselines require 27.5--104.1\,min, reducing the cost of large-scale index refreshes.

\subsubsection{Index size scalability}
Figure~\ref{fig:scal} (top-right) reports the graph-only index size, excluding the stored base vectors.
All methods scale approximately linearly with $N$.
At 10M, \ourmethod requires 2560\,MB, i.e., 256\,MB per million vectors.
In comparison, ACORN-$\gamma$ requires 8325\,MB and Milvus requires 4997\,MB, while SeRF is comparable at 2818\,MB.
Filtered-HNSW is smaller (1414\,MB), consistent with its lower-degree configuration ($M{=}32$ vs.\ $M{=}64$ for \ourmethod).@

\subsubsection{Recall robustness under scale}
Figure~\ref{fig:scal} (bottom-left) reports Recall@10 at range $=10\%$ under fixed search budgets.
As $N$ increases, achieving near-perfect recall becomes harder without increasing the exploration budget.
Filtered-HNSW remains nearly flat (0.9996 to 0.9982, a $0.14\%$ drop), while \ourmethod decreases from 0.9997 to 0.9726 (a $2.71\%$ drop).
ACORN-$\gamma$ and SeRF show intermediate decreases, from 0.9949 to 0.9779 ($1.71\%$) and from 0.9910 to 0.9783 ($1.28\%$), respectively.
Despite the fixed-parameter setting, \ourmethod retains high recall (0.973) at 10M.

\subsubsection{Throughput scalability}
Figure~\ref{fig:scal} (bottom-right) shows that \ourmethod maintains stable throughput as $N$ grows, increasing slightly from 26{,}041 to 27{,}552 QPS ($+5.8\%$).
In contrast, all CPU baselines slow down with scale: ACORN-$\gamma$ decreases from 3962 to 612 QPS ($-84.5\%$), filtered-HNSW from 1552 to 1010 QPS ($-34.9\%$), SeRF from 10{,}758 to 7074 QPS ($-34.2\%$), and Milvus from 2810 to 874 QPS ($-68.9\%$).
At 10M, \ourmethod achieves 27{,}552 QPS, corresponding to $45.0\times$, $27.3\times$, $3.9\times$, and $31.5\times$ higher throughput than ACORN-$\gamma$, filtered-HNSW, SeRF, and Milvus, respectively, under the same fixed-parameter setting.

\subsubsection{Summary}
Overall, Figure~\ref{fig:scal} shows that \ourmethod scales robustly from 1M to 10M on DEEP-96 without retuning:
it exhibits the slowest build-time growth (8.4$\times$ over a 10$\times$ larger corpus), a linear and moderate graph memory footprint (256\,MB per million vectors), high recall at 10M (0.973), and stable throughput ($+5.8\%$), whereas CPU baselines degrade by 35--85\%.

%%%%%%%%%%%%%%%%%%%%%%%%%%%%%%%%%%%
\subsection{RQ5.Ablation}

In this section, we discuss the contribution of key system components and hyperparameters to validate our design choices.

\subsubsection{Impact of Local Edge}

\begin{figure}[t]
    \centering
    \includegraphics[width=1.0\linewidth]{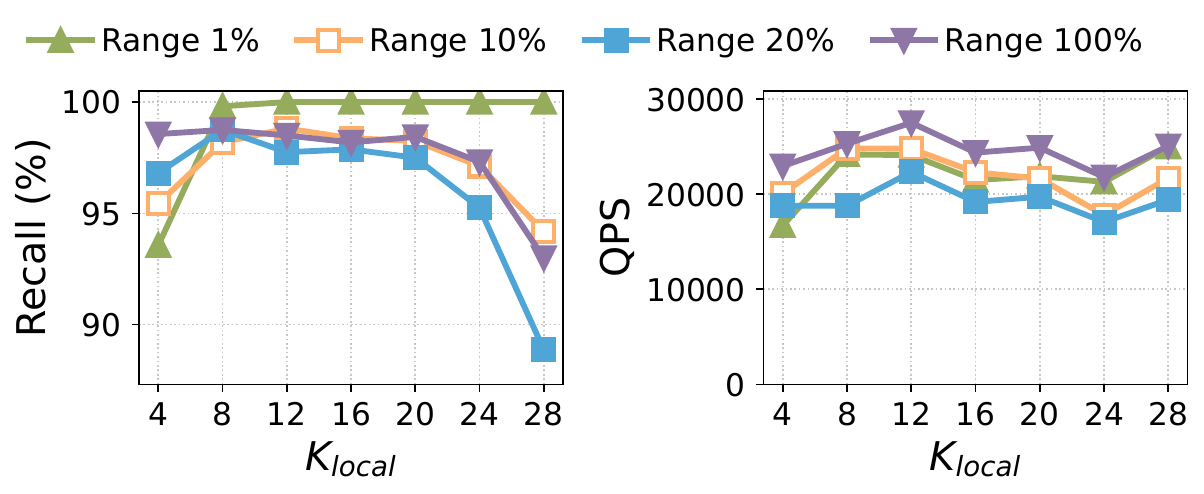}
    \caption{Test search performance on SIFT-128 with different local edges. The total degree is fixed at $K_{max}=32$, and search params are fixed at itop\_k = 128, search width = 4, maxIter = 50.}
    \label{fig:local}
\end{figure}

A core premise of \ourmethod is the \textit{Edge Redundancy Hypothesis}: a small number of local edges suffices to preserve local density, and the remaining degree budget should be allocated to remote edges for global connectivity. 
To verify this, we vary the local degree $K_{local}$ from 4 to 28 while maintaining a fixed total degree $K_{max} = 32$. Consequently, increasing $K_{local}$ implicitly reduces the budget for remote edges ($K_{remote}$). Figure~\ref{fig:local} reports the Recall and QPS on SIFT-128 across four different range selectivities.

\textbf{Analysis of Diminishing Returns.}
Initially, increasing $K_{local}$ from 4 to 12 leads to a sharp improvement in Recall across all range ratios. This confirms that a minimal density is required to navigate within a bucket. 
However, beyond $K_{local}=12$, the Recall plateaued, indicating that adding more local edges yields diminishing returns for search accuracy.

\textbf{The Penalty of Over-Localization.}
Crucially, in scenarios requiring cross-bucket navigation (e.g., Range Ratio 0.1, 0.2, and 1.0), setting an excessively high local degree (e.g., $K_{local}=28$) causes a \textbf{Recall Collapse}. 
For instance, at Range Ratio 1.0 (Global Search), Recall drops significantly from $\sim99\%$ (at $K_{local}=12$) to $\sim93\%$ (at $K_{local}=28$).
This phenomenon occurs because consuming the entire degree budget for local connections leaves insufficient capacity for remote edges. The graph becomes fragmented into isolated buckets ("islands"), making it difficult for the search to jump out of local optima and explore the global manifold.

\textbf{Conclusion.}
The experiments confirm that allocating $K_{local} \approx 12 \sim 16$ (roughly $40\%-50\%$ of the total budget) represents the optimal \emph{Sweet Spot}, balancing local precision with global navigability.

\subsubsection{Impact of Search Hyperparameters}
\label{subsubsec:ablation_params}

\begin{figure}
    \centering
    \includegraphics[width=1.0\linewidth]{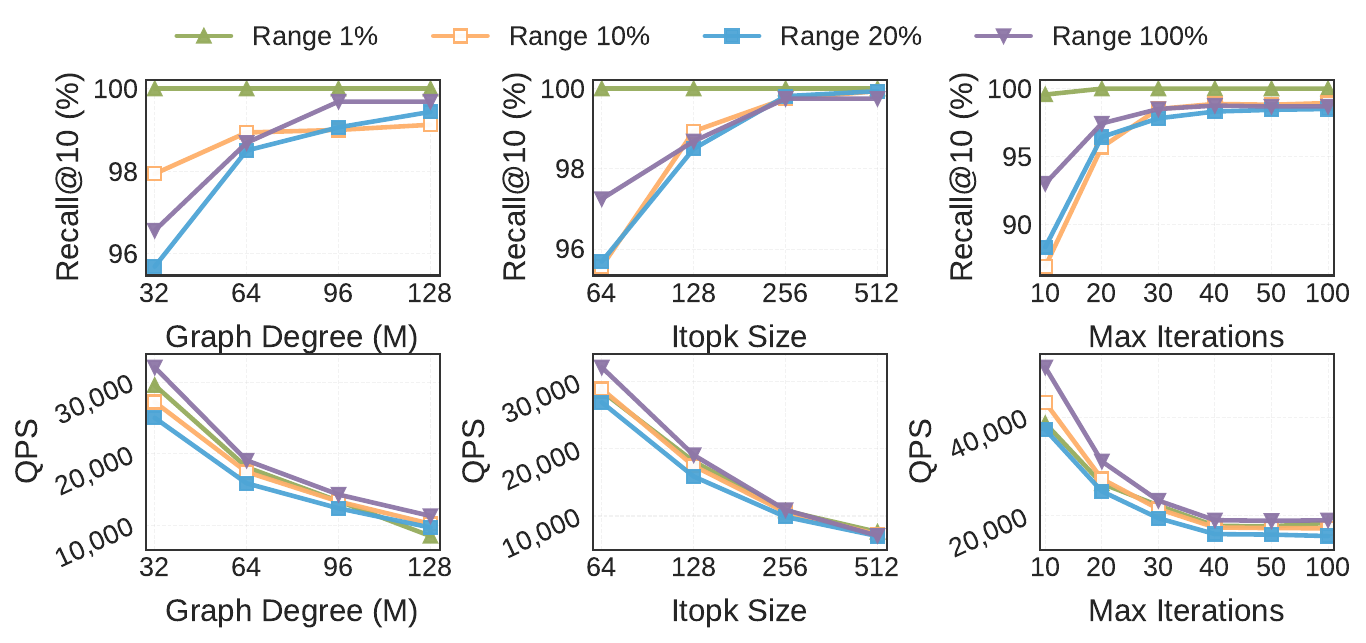}
    \caption{\textbf{Parameter sweep results for \ourmethod on DEEP-96. We vary graph degree, itopk size, and max iterations, and report recall and QPS across range ratios (1\%, 10\%, 20\%, 100\%).}}
    \label{fig:parameter}
\end{figure}

To understand the operational trade-offs of \ourmethod, we analyze the sensitivity of search performance to three critical hyperparameters: Graph Degree ($M$), Candidate Queue Size ($itopk$), and Search Depth ($max\_iter$). Figure~\ref{fig:parameter} illustrates the Recall and QPS trends on SIFT-128.

\textbf{Effect of Graph Degree ($M$).}
Increasing the graph degree (i.e., the number of neighbors stored per node) improves the connectivity of the graph, thereby enhancing Recall, especially for large-range queries (e.g., 100\% range). However, this comes at a steep cost in throughput. As $M$ increases from 32 to 128, QPS drops by nearly $3\times$. This is because scanning longer adjacency lists increases memory bandwidth pressure and cache misses on the GPU. Notably, for highly selective queries (1\% Range), a small degree ($M=32$) is sufficient to achieve 100\% recall, as the search space is naturally constrained.

\textbf{Effect of Candidate Queue Size ($itopk$).}
The $itopk$ parameter controls the width of the beam search. The results indicate a sharp threshold behavior: increasing $itopk$ from 64 to 128 yields a massive gain in Recall (from $\sim$95\% to $\sim$99\%) with a manageable drop in QPS. However, pushing $itopk$ beyond 256 offers diminishing returns in accuracy while linearly increasing the computational overhead of distance calculations. Thus, $itopk=128 \sim 256$ represents the optimal operating point.

\textbf{Effect of Search Iterations ($max\_iter$).}
This parameter limits the number of greedy hops. The curves show that \ourmethod converges rapidly. The Recall saturates around $30 \sim 40$ iterations across all range selectivities. Increasing iterations beyond 50 provides negligible accuracy gains but continues to degrade QPS. This fast convergence validates the efficiency of our Hybrid Topology, which allows the search to navigate quickly to the target neighborhood via remote edges.

\subsubsection{Impact of Freshness Bias ($\alpha$)}
\label{subsubsec:ablation_trick}

\begin{figure}[t]
    \centering
    \includegraphics[width=1.0\linewidth]{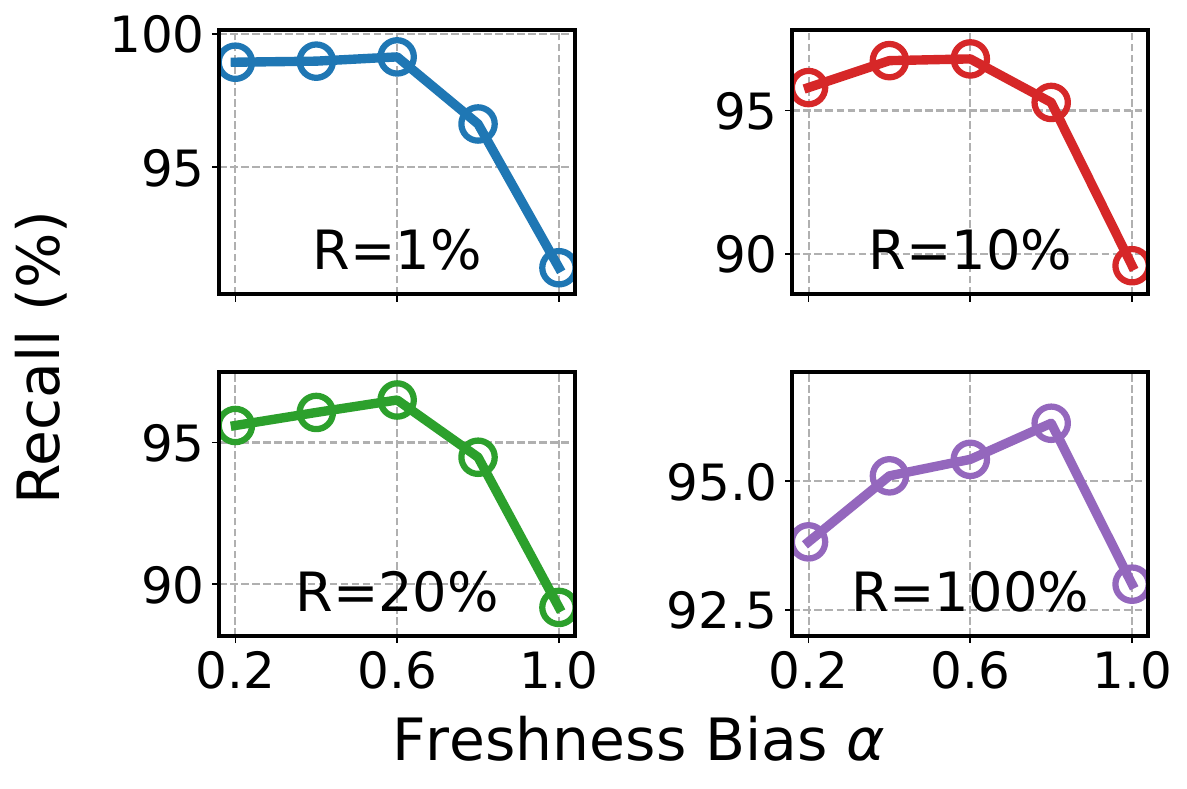}
    \caption{\textbf{Impact of Freshness Bias Coefficient ($\alpha$) on Recall.} The test is performed on SIFT-128 with dynamic insertion. Search parameters are fixed: $itop\_K=128$, $search\_width=4$. $\alpha=1.0$ denotes standard pruning without bias.}
    \label{fig:trick}
\end{figure}

In Section~\ref{subsec:insertion}, we proposed the \textit{Freshness-Biased Pruning} strategy, controlled by coefficient $\alpha$, to prevent new nodes from being rejected by established neighbors. 
Figure~\ref{fig:trick} evaluates the impact of $\alpha$ on search quality (Recall) using the SIFT-128 dataset under dynamic insertion. The value $\alpha=1.0$ represents the baseline standard pruning (no bias), while $\alpha < 1.0$ artificially increases the priority of new nodes.

\textbf{Validation of "Newcomer Isolation".}
The results show a clear performance degradation as $\alpha$ approaches 1.0. 
Notably, for selective filters (e.g., Range 1\% and 20\%), the Recall drops sharply when $\alpha=1.0$. 
This confirms our hypothesis: without the freshness bias, the "entrenched" edges in the fixed-degree graph form a barrier, leaving newly inserted nodes under-connected (isolated) and thus unreachable during search.

\textbf{Optimal Bias Range.}
Setting $\alpha$ between $0.4$ and $0.6$ yields the optimal Recall across all range ratios. This range effectively balances the trade-off between accommodating new nodes and preserving existing high-quality connections. Consequently, we adopt $\alpha=0.6$ (or 0.4) as the default setting in \ourmethod to ensure topological robustness during high-throughput ingestion.

\section{Related work}

\subsection{Graph-based ANN Indices}

% CPU 侧图算法
Graph-based indices currently represent the state-of-the-art for ANNS due to their superior trade-off between recall and latency. 

The development of graph-based indices began with the introduction of NSW (Navigable Small World) graph~\cite{malkov2014approximate}, which leverage small-world network properties to enable efficient greedy routing. Building on this  , HNSW (Hierarchical Navigable Small World)~\cite{malkov2018efficient} was proposed to further enhance performance by incorporating a hierarchical structure. It utilizes multi-layer navigation to facilitate long-range jumps in upper layers, allowing the search to quickly converge to the target neighborhood and significantly improving query efficiency. Another graph-based algorithm is NSG~\cite{zhu2021understanding}, which optimized the graph topology by approximating the Monotonic Relative Neighborhood Graph (MRNG) to reduce the average degree while maintaining connectivity.

% GPU侧图算法
With the rapid evolution of modern GPUs and heterogeneous computing platforms, researchers have increasingly explored GPU-accelerated ANN search to exploit massive parallelism and high memory bandwidth~\cite{zhao2025ame}. \textbf{SONG}~\cite{9101583} pioneered this direction by implementing graph traversal on GPUs using open-addressing hash tables and bounded priority queues, achieving 10--180$\times$ speedup over a single-threaded CPU baseline. Subsequent works, such as \textbf{GGNN}~\cite{9739943} and \textbf{GANNS}~\cite{9835618}, further moved the ANN pipeline onto the GPU. \textbf{GGNN} proposed a GPU-friendly architecture optimized for large-batch queries, demonstrating strong performance under high concurrency. \textbf{GANNS} introduced a parallel proximity graph construction algorithm, enabling efficient end-to-end indexing on GPU hardware.
%% Accelerating Graph Indexing for ANNS on Modern CPUs

% GPU侧最吊图算法
Most recently, \textbf{CAGRA}~\cite{10597683} has established the new state-of-the-art for GPU-based ANNS. It constructs a highly optimized $k$-NN graph and persists it in a static Compressed Sparse Row (CSR) format. This rigid memory layout allows CAGRA to maximize memory bandwidth and parallelism, significantly outperforming previous methods like SONG and GANNS. However, this performance comes at the cost of flexibility: the static CSR structure is immutable, making CAGRA unsuitable for scenarios requiring real-time updates or dynamic range filtering without costly reconstruction.

\subsection{Hybrid Search}

Efficiently handling hybrid queries that combine vector similarity with scalar filtering remains a significant challenge. Early attempts to address this relied on loosely coupled systems that simply adjust the execution order of filtering and searching. For instance, post-filtering performs an unconstrained approximate nearest neighbor search first and filters the results subsequently. However, this approach often suffers from severe performance degradation or fails to retrieve sufficient valid results when the filter is highly selective. Conversely, pre-filtering identifies valid candidates before initiating the graph traversal. While effective for inverted indices, applying pre-filtering to graph indices naively masks out nodes, which triggers a "connectivity crisis" that fragments the graph into disconnected components and leads to exceptionally poor recall.

To preserve global navigability under arbitrary filter constraints and overcome this connectivity crisis, recent works have shifted toward directly embedding filtering logic into the graph topology. Filtered-DiskANN~\cite{10.1145/3543507.3583552}, for example, introduces a ``Stitched Graph'' that proactively connects nodes sharing similar labels during construction. While this ensures navigability, it is primarily optimized for minimizing disk I/O rather than maximizing in-memory throughput, and its complex interleaving of label-specific edges is difficult to maintain dynamically. Building on the need for efficient in-memory range filtering, SeRF~\cite{10.1145/3639324} proposes a ``Segment Graph'' structure that achieves optimal space efficiency without building separate indices. 

Pushing the versatility of hybrid queries further, other approaches focus on maintaining robust connectivity across highly dynamic or arbitrary conditions. ACORN~\cite{10.1145/3654923} proposes a predicate-agnostic framework that expands node connectivity through a topology-aware neighbor selection strategy, ensuring the induced subgraph retains small-world properties regardless of the filter applied. Similarly addressing specialized dynamic constraints, TANNS~\cite{wang2025timestamp} tackles timestamp-filtered searches by introducing a unified Timestamp Graph. It maintains connectivity during node expiration via backup neighbors and compresses the temporal index losslessly using a Historic Neighbor Tree.

\section{Conclusion}
\label{sec:conclusion}

In this paper, we presented \textbf{\ourmethod}, a GPU-native graph indexing system designed to bridge the gap between high-throughput vector search and dynamic predicate filtering. 
Existing solutions have long suffered from a dichotomy: CPU-based indices offer flexibility but lack scalability, while static GPU indices deliver speed but fail under dynamic range constraints due to the "connectivity crisis" and memory rigidity.

\ourmethod resolves these challenges through a co-design of algorithms and system architecture. 
Algorithmically, we introduced a \textbf{Bucket-Aware Hybrid Topology} that leverages edge redundancy to maintain global connectivity even under aggressive pre-filtering. 
Systematically, we designed a \textbf{Logical-Physical Decoupled Storage} (backed by append-only layouts) and a \textbf{Freshness-Biased Update Strategy}, enabling high-concurrency data ingestion without the prohibitive cost of global reconstruction.

Extensive evaluations on 100M-scale datasets demonstrate that \ourmethod achieves \textbf{orders-of-magnitude higher QPS} (up to 100$\times$) compared to state-of-the-art CPU baselines while matching the ingestion speed of static GPU builds. 
Crucially, it maintains high recall and performance stability even during massive continuous updates. 
We believe \ourmethod paves the way for next-generation real-time vector databases that require both extreme throughput and versatile query capabilities. 
Future work will explore extending this hybrid topology to support multi-attribute composite filtering and streaming deletion scenarios.

\bibliographystyle{ACM-Reference-Format}
\bibliography{sample}

\end{document}